\documentclass[prb]{revtex4}
\usepackage{graphicx}
\usepackage{dcolumn}
\usepackage{bm}
\newcommand{\be}{\begin{equation}}
\newcommand{\ee}{\end{equation}}
\newcommand{\bea}{\begin{eqnarray}}
\newcommand{\eea}{\end{eqnarray}}
\newcommand{\eps}{\varepsilon}
\newcommand{\ev}[1]{\langle#1\rangle}

\newcommand{\ihddt}{i\hbar\frac{\partial}{\partial t}}

\newcommand{\mrm}[1]{{\mathrm{#1}}}

\renewcommand{\matrix}[1]{\mathbf{#1}}
\renewcommand{\vec}[1]{{\bf{#1}}}

\newcommand{\phiX}{\Phi}
\renewcommand{\phi}{\varphi}
\begin{document}
%
\title{Influence of bound and unbound electron-hole-pair populations on the
excitonic luminescence in semiconductor quantum wells}
%
\author{W.~Hoyer}
\email{walter.hoyer@physik.uni-marburg.de}
\author{M.~Kira}
\author{S.\,W.~Koch}
\affiliation{Department of Physics and Material Sciences Center,
          Philipps-University, Renthof 5, D-35032 Marburg, Germany}
%
\date{\today}
%
\begin{abstract}
A fully microscopic theory for the spontaneous emission from semiconductors is discussed. 
The theory is evaluated for a quantum-well system and the role of
excitonic and unbound electron-hole-pair
contributions to the emission is analyzed. Simplifying aproximations to the full
theory and their range of validity are discussed. Numerical solutions are
presented for experimentally relevant situations and it is shown that a detailed
analysis of measured spectra requires the knowledge of both
absorption and emission under identical conditions.
\end{abstract}
\maketitle
%

\section{Introduction}
\label{sec:intro}

In this paper, we present a fully 
quantum mechanical analysis of
spontaneous light emission from a Coulomb interacting
carrier system in a semiconductor quantum well (QW).  
We are particularly interested to analyze the influence
of excitonic and of unbound electron-hole-pair populations
on the photo luminescence (PL) emitted at the spectral position of
the exciton resonances.

Extensive experimental research in the past few decades has shown  
that one observes excitonic resonances in the PL of a QW for 
different material systems and for 
a very broad range of excitation conditions ranging
from cryogenic (about 4~K) 
all the way to room temperatures.\cite{Stanley:91,Leroux:99}
A PL resonance at the lowest $1s$-exciton state has also been
observed even when the exciton ionizaton density is exceeded. For example, 
experiments with GaAs-type QW systems at 4~K lattice temperature 
show a clear excitonic resonance at the $1s$ position for a density range
$10^{8} {\rm cm}^{-2}$ to $5 \times 10^{10} {\rm cm}^{-2}$\cite{Khitrova:99}; 
here, the lower density is only limited by the sensitivity in 
detection while the
upper value is already beyond the ionization density.

The build-up of excitonic PL is often investigated
in a configuration where the semiconductor is excited nonresonantly using ultrafast 
laser pulses that generate carriers into the high-energy states of the bands. 
Thus, all coherences vanish almost instantaneously and 
the direct generation of $1s$ excitons can be excluded. 
Using this scheme, it has been observed\cite{Kumar:96,Hayes:02} 
that an excitonic $1s$ PL resonance builds up on a subpicosecond timescale.
Furthermore, excitonic PL is observed 
longer than several nanoseconds, which also implies that electron-hole pairs 
exist in the system for a very long time.

In order to explain some of the observe phenomena, Ref. \cite{Kira:98}
developed a simple microscopic model for the semiconductor PL
showing that also an electron-hole plasma, i.e. a system with
nonvanishing electron and hole densities but no excitons, can 
produce PL spectra that peak at the exciton resonances. 
This prediction releases the previously assumed strict relation between 
the existence of exciton populations and 
the appearance of excitonic PL such that one can explain
why excitonic PL is also observed under conditions where the presence
of real exciton populations seems unreasonable.  

In order to present a more complete
analysis, we show in this paper how one can analyze self-consistently
the electron-hole plasma and exciton-population effects in
the QW PL. We show that in order to perform a 
reliable analysis, one has to evaluate not only the PL
but also the corresponding absorption spectra 
recorded under equivalent conditions. Ideally, one should also
measure the induced terahertz (THz) response\cite{Timusk:78,Groeneveld:94,Cerne:96,Kira:01,Kaindl:03,Galbraith:05}.

Based on our microscopic analysis,
we determine criteria under which exciton population 
effects are observed and when electron-hole plasma contributions are dominant.
We also discuss the Coulomb-induced nonlinearities 
observable in PL for elevated densities. For this purpose,
one clearly has to
perform a quantitative analysis where the microscopic scattering,
electron-hole plasma, and possible exciton populations
have to be included microscopically. 
The developed analysis allows
us to determine systematic criteria for a measured combination 
of absorption and PL
to judge how much of a given PL spectrum can be 
attributed to excitonic and/or electron-hole plasma emission.

The discussion in this paper is organized as follows. We first present 
our general quantum theory for PL from coexisting electron-hole plasma 
end exciton populations. We show how this approach, 
with microscopic Coulomb scattering,
can be formulated to compute the PL from quasi-stationary
semiconductor populations consisting of a given exciton 
and electron-hole plasma population. We evaluate our theory
numerically for a two-band QW model and present a detailed
analysis of the exciton and electron-hole-plasma contibutions
to the luminescence. We discuss different levels of
simplifying approximations and emphasize the importance
of the frequency and density dependent scattering. 
Finally, we summarize our results and use them to analyze
experimental data. In the Appendices, we present technical
details regarding the microscopic description of excitons and
the density and frequency dependent scattering calculations.

\section{Microscopic theory of excitonic luminescence}
\label{sec:theory}

We formulate our microscopic theory for the incoherent emission
of a planar quantum well. In our calculations, we concentrate
on PL resulting from recombinations involving only two bands, i.e.
the energetically lowest conduction and the heavy-hole
band. Multi-band effects can in principle be included in the theory,
but they complicate the notation and do not contribute
to the essence of our analysis.
In the numerical evaluations, we use typical GaAs QW material 
parameters and choose the QW confinement such that the $2s$ resonance
is 6\,meV above the $1s$ state as observed in recent experiments.\cite{Chatterjee:04,Szczytko:04,Hoyer:05c,Szczytko:05} 
For semiconductors, the microscopic description of 
carrier excitations can be obtained with the help of 
the fermionic operators $a_{c,{\bf k}_\|}$ and $a^\dagger_{c,{\bf k}_\|}$
for conduction-band electrons and $a_{v,{\bf k}_\|}$ 
and $a^\dagger_{v,{\bf k}_\|}$
for valence-band electrons having momentum ${\bf k}_\|$ in the QW plane. 
The quantized light field is described by bosonic operators
$B_{{\bf q}_\|,q_\perp}$ and $B^\dagger_{{\bf q}_\|,q_\perp}$
related to free space eigen modes with momentum
${\bf q} = ({\bf q}_\|,q_\perp)$ that is divided into the ${\bf q}_\|$
component in the QW plane and $q_\perp$ perpendicular to the QW.
The full system Hamiltonian of the interacting carrier-photon system is discussed
e.g.~in Refs.~\onlinecite{Haug:04,Kira:99} and the relating microscopic 
PL dynamics is derived in 
Refs.~\onlinecite{Kira:97a,Kira:98,Kira:99,Kira:01,Hoyer:03,Kira:05,Kira:06}.
Here, we therfore present only a brief summary of
the elements important for our analysis of excitonic PL.

Since we want to compute PL 
from incoherent quasiparticle excitations, we concentrate on
conditions where the coherent polarization and the coherent
fields $\langle B_{\bf q} \rangle$ vanish.
We also assume that the system is homogeneously excited and quasi stationary.
For such conditions, the single-particle
distributions
\begin{equation}
 f^{e}_{{\bf k}_\|} \equiv 
 \langle a^\dagger_{c,{\bf k}_\|,c} a_{c,{\bf k}_\|} \rangle,
 \;\;\;\;
 f^{h}_{{\bf k}_\|} \equiv 
 \langle a_{v,{\bf k}_\|} a^\dagger_{v,{\bf k}_\|}\rangle
\label{eq:densities},
\end{equation}
for the electrons and holes can be taken as constant.

The relevant two-particle correlations in the incoherent regime are excitonic
and pure intraband carrier correlations.\cite{Kira:01,Hoyer:03} For the study of PL,
we only need to evaluate the excitonic correlations defined as
\begin{eqnarray} 
&&c_{X}^{{\bf q}_\|,{\bf k}'_\|,{\bf k}_\|} 
 \equiv \Delta 
\langle 
a^{\dagger}_{c,{\bf k}_\|} a^{\dagger}_{v,{\bf k}'_\|}  
a_{c,{\bf k}'_\|+{\bf q}_\|}  a_{v,{\bf k}_\|-{\bf q}_\|} \rangle
\nonumber\\
&&
\;\;\;\;\;\;\;\;\;\;
=
\langle 
a^{\dagger}_{c,{\bf k}_\|} a^{\dagger}_{v,{\bf k}'_\|}  
a_{c,{\bf k}'_\|+{\bf q}_\|}  a_{v,{\bf k}_\|-{\bf q}_\|} \rangle
- 
\langle 
a^{\dagger}_{c,{\bf k}_\|} a^{\dagger}_{v,{\bf k}'_\|}  
a_{c,{\bf k}'_\|+{\bf q}_\|}  a_{v,{\bf k}_\|-{\bf q}_\|} \rangle_{\rm S}
\label{eq:two_ple},
\end{eqnarray} 
where the subtracted single-particle contribution $\langle a^{\dagger}_{\lambda} a^{\dagger}_{\nu}  
a_{\nu'}  a_{\lambda'} \rangle_{\rm S}$ is obtained by using the
usual Hartree-Fock approximation
$\langle a^{\dagger}_{\lambda} a^{\dagger}_{\nu}  
   a_{\nu'}  a_{\lambda'} \rangle_{\rm S}
   \equiv
   \langle a^{\dagger}_{\lambda} a_{\lambda'} \rangle
   \langle a^{\dagger}_{\nu} a_{\nu'} \rangle
   -
    \langle a^{\dagger}_{\lambda} a_{\nu'} \rangle
   \langle a^{\dagger}_{\nu} a_{\lambda'} \rangle$.

The incoherent photon correlations and the microsocpic source for the
PL are obtained from the photon-assisted electron-hole 
recombination and the photon-number-like correlations
\begin{eqnarray}
&&\Pi_{{\bf k}_\|,{\bf q}_\|,q_\perp} \equiv
\Delta \langle  B^{\dagger}_{{\bf q}_\|,q_\perp} 
  a_{v,{\bf k}_\|}^{\dagger} a_{c,{\bf k}_\|+{\bf q}_\|} \rangle
\nonumber\\
&&
\;\;\;\;\;\;\;\;\;\;
=
\langle  B^{\dagger}_{{\bf q}_\|,q_\perp} 
  a_{v,{\bf k}_\|}^{\dagger} a_{c,{\bf k}_\|+{\bf q}_\|} \rangle
  - 
  \langle  B^{\dagger}_{{\bf q}_\|,q_\perp} \rangle 
  \langle a_{v,{\bf k}_\|}^{\dagger} a_{c,{\bf k}_\|+{\bf q}_\|} \rangle,
\nonumber\\
&&\Delta 
   \langle B^{\dagger}_{{\bf q}_\|,q_\perp} B_{{\bf q}_\|,q'_\perp} \rangle
\equiv
\langle B^{\dagger}_{{\bf q}_\|,q_\perp} B_{{\bf q}_\|,q'_\perp} \rangle
-
\langle B^{\dagger}_{{\bf q}_\|,q_\perp} \rangle 
\langle B_{{\bf q}_\|,q'_\perp} \rangle
\label{eq:Pi_and_B*B_corr},
\end{eqnarray} 
respectively. These are obtained by subtracting the classical factorizations
from the full expectation values. 
Once these quantities are determined microscopically, we can evaluate
the semiconductor PL and study the influence of
electron-hole plasma and exciton populations.

The exciton correlation dynamics follows from
\begin{eqnarray}
  i \hbar\frac{\partial}{\partial t}
  c^{{\bf q}_\|,{\bf k}'_\|,{\bf k}_\|}_{\rm X}
&=&\epsilon^{{\bf q}_\|,{\bf k}'_\|,{\bf k}_\|} 
c^{{\bf q}_\|,{\bf k}'_\|,{\bf k}_\|}_{\rm X}
+ S^{{\bf q}_\|,{\bf k}'_\|,{\bf k}_\|}
\nonumber\\
  &+&
  \left(
     1-f^e_{{\bf k}_\|}-f^h_{{\bf k}_\|-{\bf q}_\|}
  \right)
  \sum_{{\bf l}_\|}
	V_{{\bf l}_\|-{\bf k}_\|}
   c^{{\bf q}_\|,{\bf k}'_\|,{\bf l}_\|}_{\rm X}
-
  \left(
     1-f^e_{{\bf k}'_\|+{\bf q}_\|}-f^h_{{\bf k}'_\|}
  \right) 
  \sum_{{\bf l}_\|}
      V_{{\bf l}_\|-{\bf k}'_\|}
      c^{{\bf q}_\|,{\bf l}_\|,{\bf k}_\|}_{\rm X}
\nonumber\\
&-&
 i( 1- f^e_{\vec{k}_\|}-f^h_{\vec{k}_\|-\vec{q}_\|})
     \Pi_{{\bf k}'_\|,{\bf q}_\|,\Sigma}
  -i(1-f^e_{\vec{k}'_\|-\vec{q}_\|}-f^h_{\vec{k}'_\|})
  \Pi^\star_{{\bf k}_\|-{\bf q}_\|,{\bf q}_\|,\Sigma}
\nonumber\\
&+&D^{{\bf q}_\|,{\bf k}'_\|,{\bf k}_\|}_{\rm rest}
- i T^{{\bf q}_\|,{\bf k}'_\|,{\bf k}_\|}
\label{eq:EXPcvcv},
\end{eqnarray}
where we have introduced the
renormalized electron-hole pair energy
\begin{equation}
\epsilon^{{\bf q}_\|,{\bf k}'_\|,{\bf k}_\|}
\equiv
    \epsilon^{c}_{{\bf k}_\|-{\bf q}_\|}
    - 
    \epsilon^{v}_{{\bf k}_\|}
    -
    \epsilon^{c}_{{\bf k}'_\|}
    +
   \epsilon^{v}_{{\bf k}'_\|+{\bf q}_\|} 
    -\sum_{\bf l} V_{{\bf l}_\|}
    \left[
      f^e_{{\bf k}_\|-{\bf q}_\| -{\bf l}_\|}
      +f^h_{{\bf k}_\|-{\bf l}_\|}
      -f^e_{{\bf k}'_\|-{\bf l}_\|}
      -f^h_{{\bf k}'_\|+{\bf q}_\|-{\bf l}_\|}
    \right]
\label{eq:E_exciton}.
\end{equation}
Furthermore, we have defined a collective photon assisted electron-hole
recombination term
\begin{eqnarray}
&&
   \Pi_{{\bf k}_\|,{\bf q}_\|,\Sigma}
   \equiv
   \Delta \langle  B^{\dagger}_{{\bf q}_\|,\Sigma} 
  a_{v,{\bf k}_\|}^{\dagger} a_{c,{\bf k}_\|+{\bf q}_\|} \rangle,
\label{eq:Pi_sigma}\\
   &&
   B_{{\bf q}_\|,\Sigma} \equiv \sum_{q_\perp} 
   {\cal F}_{{\bf q},q_\perp} B_{{\bf q},q_\perp}
\label{eq:B_sigma},
\end{eqnarray}
where ${\cal F}_{{\bf q}} \equiv
d_{vc} \sqrt{\frac{\hbar \omega_{\bf q}}{2 \epsilon_0}} u_{\bf q}$ 
determines the strength of the light-matter interaction
via the dipole-matrix element $d_{vc}$ and
the mode function $u_{\bf q}$ at the QW position.
The square-root term is the so-called vacuum field amplitude containing
the light frequency $\omega_{\bf q}$. 

The single-particle source of Eq.~(\ref{eq:EXPcvcv}) originates from the 
Coulomb interaction and has the explicit form 
\begin{eqnarray}
  S^{{\bf q}_\|,{\bf k}'_\|,{\bf k}_\|}
  &=&
  V_{{\bf k}'_\|+{\bf q}_\| - {\bf k}_\|} 
  \left[ 
    (1-f^e_{{\bf k}_\|}) (1-f^h_{{\bf k}_\|-{\bf q}_\|})
    f^e_{{\bf k}'_\|+{\bf q}_\|} f^h_{{\bf k}'_\|}
    -
    f^e_{{\bf k}_\|} f^h_{{\bf k}_\|-{\bf q}_\|}
    (1-f^e_{{\bf k}'_\|+{\bf q}_\|}) (1-f^h_{{\bf k}'_\|})
  \right]
\label{eq:S_X}.
\end{eqnarray}
This expression describes the Coulomb-induced scattering contributions
which develop the electron-hole plasma into a correlated state
even when no correlations are present initially.

Equation (\ref{eq:EXPcvcv}) also contains several
Coulomb sums. Here, we have explicitly presented only
those terms that contain the 
phase-space filling factor $(1-f^e-f^h)$ 
since they describe
the attractive interaction between electrons and holes
allowing them to become truly bound electron-hole pairs,
i.e.~{\it incoherent excitons}. These Coulomb sums
are referred to as the {\it main-sum contributions}.\cite{Hoyer:03}
The other Coulomb sums, denoted symbolically as 
$D_{\rm rest}$, consist of terms of the form 
$(f^\lambda - f^\nu) \sum V \; c$. Since $(f^\lambda - f^\nu)$ is much smaller
than $(1-f^e-f^h)$ for low enough densities, $D_{\rm rest}$
has a much smaller effect on the exciton correlations than the main-sum
terms for conditions where bound excitons still exist.\cite{Hoyer:03} 
Thus, we may apply the main-sum approximation where
$D_{\rm rest}$ is omitted when we compute the excitonic
PL. Besides the two-particle correlations, 
Eq.~(\ref{eq:EXPcvcv}) also contains
three-particle correlation terms symbolized by  
$T^{{\bf q},{\bf k}',{\bf k}}$. The description of these terms is presented
in Appendix B. 

The incoherent carrier quantities enter as sources to the
{\it semiconductor luminescence equations} \cite{Kira:97a,Kira:99,Kira:04,Kira:05}
\begin{eqnarray}
   &&\frac{\partial}{\partial t} 
   \Delta \langle B^{\dagger}_{{\bf q}_\|,q_\perp} 
   B_{{\bf q}_\|,q'_\perp} \rangle 
   = 
   i \left( \omega_{q} - \omega_{q'} \right) 
   \Delta \langle B^{\dagger}_{{\bf q}_\|,q_\perp}
   B_{{\bf q}_\|,q'_\perp} \rangle
   +{\textstyle \sum_{{\bf k}}}
   \left[
     {\cal{F}}_{q} \Pi^\star_{{\bf k}_\|,{\bf q}_\|,q'_\perp}
     +
     {\cal{F}}^{\star}_{q'} 
     \Pi_{{\bf k}_\|,{\bf q}_\|,q_\perp}
   \right]
\label{eq:SLE_b*b}
\\
 && 
  \frac{\partial}{\partial t}
  \Pi_{{\bf k}_\|,{\bf q}_\|,q_\perp}
  = 
  \frac{1}{i\hbar}
  \left[
    \left(
    \tilde{\epsilon}_{{\bf k}_\|,{\bf q}_\|}
    - \hbar\omega_{\bf q}
    \right)
    \Pi_{{\bf k}_\|,{\bf q}_\|,q_\perp}
    -
    \left[1-f^e_{{\bf k}_\|+{\bf q}_\|} -f^h_{{\bf k}_\|}\right]
    \sum_{{\bf l}_\|} 
    V_{{\bf k}_\| - {\bf l}_\|}
    \Pi_{{\bf l}_\|,{\bf q}_\|,q_\perp}
    \right]
\nonumber\\
  &&
  +{\cal F}_q
  \left[ 
          f^e_{{\bf k}_\|+{\bf q}_\|} f^h_{{\bf k}_\|}
        +       
        \sum_{{\bf l}_\|}
        c_X^{{\bf q}_\|,{\bf k}_\|,{\bf l}_\|}
  \right] 
  -
  \left[1-f^e_{{\bf k}_\|+{\bf q}_\|} -f^h_{{\bf k}_\|}\right]
  \Delta \langle  B^{\dagger}_{{\bf q}_\|,q_\perp} 
   B_{{\bf q}_\|,\Sigma} \rangle
  - i T^{\Pi}_{{\bf k}_\|,{\bf q}_\|,q_\perp}
\label{eq:SLE_BY}.
\end{eqnarray}
Here, 
\begin{eqnarray}
   &&
  \tilde{\epsilon}_{{\bf k}_\|,{\bf q}_\|}
   = \epsilon^c_{{\bf k}_\|+{\bf q}_\|}-\epsilon^v_{{\bf k}_\|}
   -\sum_{{\bf l}_\|} V_{{\bf k}_\|-{\bf l}_\|}
   \left[f^e_{{\bf l}_\|+{\bf q}_\|}+f^h_{{\bf l}_\|}\right]
\label{eq:E_def}
\end{eqnarray}
is the renormalized kinetic energy.
The first term in the second line of Eq.~(\ref{eq:SLE_BY}) shows 
that photon-assisted recombination processes can be seeded either 
by an electron-hole plasma population ($f^e f^h$) or an exciton correlation 
term ($c_X$). Since these spontaneous emission
sources do not depend on the photon frequency,
both plasma and exciton correlation sources contribute to the PL 
in the same spectral range. 

The second term in Eq.~(\ref{eq:SLE_BY}) shows that
the photon-assisted recombination processes 
for different momentum combinations are 
coupled via the Coulomb interaction. This renormalization 
determines the emission frequency similar to the 
classical field renormalization in the semicondcutor 
Bloch equations \cite{Lindberg:88,Haug:04,Kira:04b,Kira:06}.
Since this Coulomb correlation
effect is not at all related to the 
specific spontaneous emission source,
even the existence of bare electron and hole densities leads to
emission of light at the bound exciton frequencies. 
When the QW is not inside a cavity, the stimulated feedback term,
$\Delta \langle B^\dagger B_\Sigma \rangle$, in Eq.~(\ref{eq:SLE_BY}) is negligible.
The explicit form of the three-particle scattering term,
$T^{\Pi}$, is discussed in Appendix B.

\subsection{Excitons vs.~correlated electron-hole plasma}
\label{sec:def_X}

As a very broad definition, an exciton is a quasi-particle that
consist of a bound electron-hole pair. This interpretation
is seemingly simple and straightforward. However, once one seriously
considers the fact that an exciton has a fermionic substructure,
as is done in Eq.~(\ref{eq:EXPcvcv}),
one realizes that one can neither define a pure exciton operator
nor a rigorous number operator for excitons in a fermionic
many-body system.\cite{Usui:60} As a working definition, we
speak of an exciton in a many-particle system 
if at least one electron depends on a hole in a correlated manner,
where the correlation is in the relative motion coordinate. The 
pair wavefunction is determined 
from the Wannier equation that is analogous to the generalized
Schr\"{o}dinger equation for the hydrogen. The
bound state solutions define the bound excitons while the ionized
states define ionized excitons.

The Wannier equation in momentum space is equivalent to the Fourier transform
of the Hydrogen equation, but can contain additional energy renormalizations
and a phase-space filling factor if a finite amount of carriers is present.\cite{Haug:04}
As shown in Appendix A, it is even possible to incorporate carrier
scattering in the Wannier equation and thus obtain excitonic quasi-particles
including Coulomb scattering effects. In the presence of carriers, both the 
phase-space filling factor and the microscopic scattering make the
corresponding eigenvalue problem non-Hermitian, such that a left-handed
and a right-handed basis set ($\phiX_{\nu,\vec{q}_{\|}}^{L/R}$) must be introduced, see
Appendix A. There, we show that the definition
of true excitons can be mathematically formulated by using the
transformations
\begin{eqnarray}
      \Delta N_{\lambda,\nu}({\bf q}_\|) 
        &=&
	\sum_{{\bf k}_\|,{\bf k}'_\|} 
	\phiX^L_{\lambda,\vec{q}_{\|}}({\bf k}_\|) \left(\phiX^L_{\nu,\vec{q}_{\|}}({\bf k}'_\|) \right)^\star
	c^{{\bf q}_\|,{\bf k}'_\|-{\bf q}_h,{\bf k}_\|+{\bf q}_e}_{\rm X},
\\
c^{{\bf q}_\|,{\bf k}'_\|,{\bf k}_\|}_{\rm X}
&=&\sum_{\lambda,\nu}
	\left( \phiX^R_{\lambda,\vec{q}_{\|}}({\bf k}_\|-{\bf q}_e) \right)^\star
	\phiX^R_{\nu,\vec{q}_{\|}}({\bf k}'_\|+{\bf q}_h)
	\Delta N_{\lambda,\nu}({\bf q}_\|),
\label{eq:XX_transform}
\end{eqnarray}
where the excitonic center-of-mass momentum has been split symmetrically via
\begin{eqnarray}
{\bf q}_{e} = \frac{m_{e}}{m_e+m_h} {\bf q}_\|,
\;\;\;\;
{\bf q}_{h} = \frac{m_{h}}{m_e+m_h} {\bf q}_\|
\label{eq:q_eh}.
\end{eqnarray}
In particular, the diagonal elements,
\begin{eqnarray}
&&\Delta N_\lambda ({\bf q}_\|) \equiv
\Delta N_{\lambda,\lambda}({\bf q}_\|),
\label{eq:X_distribution}
\end{eqnarray}
define the momentum distribution for excitons in state $\lambda$.

The physical relevance of these quantities can be verified
by analyzing THz absorption which is shown\cite{Kira:01,Kira:04}
to probe exclusively $\sum_{{\bf q}_\|} \Delta{N}_{\lambda}({\bf q}_\|)$
under incoherent conditions. More precisely, each exciton state 
can be connected with a population 
\be
\Delta{n}_{\lambda} = \frac{1}{S} \sum_{{\bf q}_\|} \Delta{N}_{\lambda}({\bf q}_\|) 
\ee
and a $1s$-to-$2p$ resonance in the linear THz absorption 
shows unambiguously that the $1s$ state is initially populated.
In full analogy to the optical transitions in atomic systems, 
the strength of the resonance is directly related to the 
population difference between the initial $\Delta n_{1s}$
and the final $\Delta n_{2p}$ states. Thus, THz spectroscopy
shows that $\Delta n_\beta$ defines the exciton density
of a given state in the same way as one can count atoms occupying
different atomic states by using optical spectroscopy.

In general, exciton correlations can also exhibit nondiagonal
$\lambda \neq \nu$ contributions in $\Delta N_{\lambda,\nu}$. 
As shown in Ref. \cite{Hoyer:03}, these off-diagonal parts 
build up on a pico-second timescale and can be identified with a 
correlated plasma while the diagonal exciton populations are formed
much more slowly. In the present publication we include microscopic 
Coulomb scattering beyond the constant dephasing and we include both
diagonal and non-diagonal exciton correlations.

\subsection{Incoherent quasiparticle states}
\label{sec:absoprtion}

As already argued in \cite{Chatterjee:04}, it is essential
to analyze the experimental PL together
with absorption spectra obtained under identical conditions. 
Since many experiments measure PL in the normal
direction, we consider this configuration here. Using
our formalism, we establish a 
connection between the quasi-particle state and the measured spectra
in order to determine the conditions and excitation parameters
where excitonic populations significantly influence PL experiments.

When a weak classical optical pulse, $E(t)$, is applied to probe the
incoherent excitation state of a semiconductor, it
does not alter electron, hole, or exciton populations. The probe pulse 
interacts with the system, merely by inducing the optical (probe) polarization 
$P_{{\bf k}_\|} \equiv
\langle a^\dagger_{v,{\bf k}_\|}a_{c,{\bf k}_\|} \rangle$
in the system. Thus, we need to solve the polarization
dynamics from the semiconductor Bloch equations \cite{Lindberg:88,Haug:04}
\begin{eqnarray}
  &&i \hbar \frac{\partial}{\partial t} 
  P_{{\bf k}_\|}
  = 
  \tilde{\epsilon}_{{\bf k}_\|}
  P_{{\bf k}_\|} 
- \left[1-f^e_{{\bf k}_\|}-f^h_{{\bf k}_\|}\right] 
  \sum_{{\bf k}'_\|} V_{{\bf k}'_\|-{\bf k}_\|}
  P_{{\bf k}'_\|}
\nonumber\\
&&
\;\;\;\;\;\;\;\;\;\;
- \left[1-f^e_{{\bf k}_\|}-f^h_{{\bf k}_\|}\right] 
  d_{c,v} E(t)
  - i \Gamma_{{\bf k}_\|}
 \label{eq:SBE-Pol}
\end{eqnarray}
which contains two-particle Coulomb- and phonon-induced scattering terms in
$\Gamma_{{\bf k}_\|}$. Performing a Fourier transformation of
Eq.~(\ref{eq:SBE-Pol}), we obtain
\bea
&&
  (\hbar \omega -   \tilde{\epsilon}_{{\bf k}_\|})
  P_{{\bf k}_\|}(\omega)
  = 
- \left[1-f^e_{{\bf k}_\|}-f^h_{{\bf k}_\|}\right] 
  \sum_{{\bf k}'_\|} V_{{\bf k}'_\|-{\bf k}_\|}
  P_{{\bf k}'_\|}(\omega)
\nonumber\\
&&
\;\;\;\;\;\;\;\;\;\;
- \left[1-f^e_{{\bf k}_\|}-f^h_{{\bf k}_\|}\right] 
  d_{c,v} E(\omega)
  - i \Gamma_{{\bf k}_\|}(\omega)
 \label{eq:SBE-Pol2},
\eea
where we describe the Coulomb-induced scattering 
\be
\Gamma_{{\bf k}_\|}(\omega)
   = 
\gamma^{\mrm{D}}_{{\bf k}_\|}(\omega) P_{{\bf k}_\|}(\omega)
-
\sum_{{\bf k}'_\|} \gamma^{\mrm{OD}}_{{\bf k}_\|,{\bf k}'_\|}(\omega) 
    P_{{\bf k}'_\|}(\omega)
\label{eq:SBE_scatt}
\ee
via the diagonal and off-diagonal excitation-induced dephasing 
matrices $\gamma^{\mrm{D}}$ and $\gamma^{\mrm{OD}}$. Their 
detailed form is given in Appendix B.

The linear absorption spectrum follows from
\begin{eqnarray}
  \alpha(\omega) \equiv 
  {\rm Im}
  \left[
    \frac{1}{\epsilon_0 {\cal S}}
    \sum_{{\bf k}_\|}  \frac{d_{{\bf k}'_\|} P_{{\bf k}'_\|}(\omega)}{E(\omega)}
    \right]
\label{eq:absorption}.
\end{eqnarray}
If the semiconductor system is in thermodynamic equilibrium,
the linear absorption, $\alpha(\omega)$ and PL spectrum,
$PL(\omega)$ are connected via\cite{Zimmermann:88} the well-known thermal
relation
\begin{equation}
   I_{PL}^{\rm TD}(\omega) \equiv 
   \frac{\alpha(\omega)}
   {e^{(\hbar \omega - \mu)/(k_B T)}-1}
\label{eq:KMS}
\end{equation}
where $\omega$ is the angular frequency of the light, $\mu$ is the chemical 
potential of electron-hole pairs, and $T$ is the lattice temperature. For
dilute densities, the Bose-Einstein distribution can be replaced by a 
Boltzmann distribution. Clearly, $I_{PL}^{\rm TD}(\omega)$ provides
a well-defined reference point under all excitation conditions.
In fact, the relative contribution of the electron-hole plasma to
the excitonic PL can be
identified by quantifying how strongly the actually measured PL
deviates from the thermodynamic limit.

Under quasistationary conditions, the steady-state PL 
emitted in normal direction, ${\bf q}_\| =0$, is obtained from
Eq.~(\ref{eq:SLE_b*b})--(\ref{eq:SLE_BY}) as
\begin{eqnarray}
   &&
  I_{PL}(\omega = c|{\bf q}|)
  \equiv
  \frac{\partial}{\partial t} 
   \Delta \langle B^{\dagger}_{0,q_\perp} 
   B_{0,q_\perp} \rangle 
   = 
   {\rm Re}\left[
     \sum_{{\bf k}_\|}
	 {\cal{F}}^{\star}_{q'} 
	 \Pi_{0,{\bf q}_\|,q_\perp}
   \right]
\label{eq:SLE_b*b_steady}
\\
 && 
  \frac{\partial}{\partial t}
  \Pi_{{\bf k}_\|,0,q_\perp}
  = 
  \frac{1}{i\hbar}
  \left[
    \left(
    \tilde{\epsilon}_{{\bf k}_\|,{\bf q}_\|}
    - \hbar\omega_{\bf q}
    \right)
    \Pi_{{\bf k}_\|,0,q_\perp}
    -
    \left[1-f^e_{{\bf k}_\|} -f^h_{{\bf k}_\|}\right]
    \sum_{{\bf l}_\|} 
    V_{{\bf k}_\| - {\bf l}_\|}
    \Pi_{{\bf l}_\|,0,q_\perp}
    \right]
\nonumber\\
  &&
\;\;\;\;\;\;\;\;\;\;
\;\;\;\;\;\;\;\;\;\;
  +{\cal F}_q
  \left[ 
          f^e_{{\bf k}_\|} f^h_{{\bf k}_\|}
        +       
        \sum_{{\bf l}_\|}
        c_X^{0,{\bf k}_\|,{\bf l}_\|}
  \right] 
- i
  T^{\Pi}_{{\bf k}_\|,0,q_\perp}
\label{eq:SLE_BY_steady}.
\end{eqnarray}
Here, we have omitted the stimulated contributions since they are insignificant
for bare QW systems. As shown in App.~\ref{app:scat}, the three-particle
scattering terms can be cast into a form similar to the 
polarization scattering in Eq.~(\ref{eq:SBE_scatt}), i.e.
\begin{eqnarray}
&&  
 T^{\Pi}_{{\bf k}_\|,0,q_\perp}
 \equiv
\gamma^{\mrm{D}}_{{\bf k}_\|}(\omega_{q_\perp}) \Pi_{{\bf k}_\|,0,q_\perp}
-
\sum_{{\bf k}'_\|} \gamma^{\mrm{OD}}_{{\bf k}_\|,{\bf k}'_\|}(\omega) 
\Pi_{{\bf k}'_\|,0,q_\perp}.
\label{eq:SLE_scatt}
\end{eqnarray}
Thus, the semiconductor Bloch and luminescence
equations exhibit the same generic structure. However, as the major difference,
the semiconductor PL has as its incoherent source the combination of
carrier densities $f^e f^h$ and exciton correlations while the
optical polarization is driven by the coherent field $E(\omega)$.

The incoherent quasi-particle correlations
enter Eq.~(\ref{eq:SLE_BY_steady})
via $c_X$. To separate the exciton-population and
correlated electron-hole plasma contributions, we introduce
\begin{eqnarray}
&&c^{0,{\bf k}'_\|,{\bf k}_\|}_{\rm X}
=\sum_{\lambda}
	\left( \phiX^R_{\lambda,\vec{q}_{\|}}({\bf k}_\|) \right)^\star
	       \phiX^R_{\lambda,\vec{q}_{\|}}({\bf k}'_\|)
	\Delta N_{\lambda,\lambda}(0) +
 c^{0,{\bf k}'_\|,{\bf k}_\|}_{\rm plasma}
\label{eq:X_vs_plasma},
\end{eqnarray}
where the steady-state form of the
correlated plasma part is obtained from  Eq.~(\ref{eq:EXPcvcv})
\begin{eqnarray}
  &&i \hbar\frac{\partial}{\partial t}
  c^{0,{\bf k}'_\|,{\bf k}_\|}_{\rm plasma}
=\epsilon^{0,{\bf k}'_\|,{\bf k}_\|} 
c^{0,{\bf k}'_\|,{\bf k}_\|}_{\rm X}
\nonumber\\
&&
\;\;\;\;\;\;\;\;\;\;
+ 
  V_{{\bf k}'_\|- {\bf k}_\|} 
  \left[ 
    (1-f^e_{{\bf k}_\|}) (1-f^h_{{\bf k}_\|})
    f^e_{{\bf k}'_\|} f^h_{{\bf k}'_\|}
    -
    f^e_{{\bf k}_\|} f^h_{{\bf k}_\|}
    (1-f^e_{{\bf k}'_\|}) (1-f^h_{{\bf k}'_\|})
  \right]
\nonumber\\
  &&
\;\;\;\;\;\;\;\;\;\;
+
  \left(
     1-f^e_{{\bf k}_\|}-f^h_{{\bf k}_\|}
  \right)
  \sum_{{\bf l}_\|}
	V_{{\bf l}_\|-{\bf k}_\|}
   c^{0,{\bf k}'_\|,{\bf l}_\|}_{\rm plasma}
-
  \left(
     1-f^e_{{\bf k}'_\|}-f^h_{{\bf k}'_\|}
  \right) 
  \sum_{{\bf l}_\|}
      V_{{\bf l}_\|-{\bf k}'_\|}
      c^{0,{\bf l}_\|,{\bf k}_\|}_{\rm plasma}
\nonumber\\
&&
\;\;\;\;\;\;\;\;\;\;
- i
  \sum_{{\bf l}_\|}
  \left[\gamma^X_{{\bf k}_\|,{\bf l}_\|}\right]^\star
   c^{0,{\bf k}'_\|,{\bf l}_\|}_{\rm plasma}
- i
  \sum_{{\bf l}_\|}
  \gamma^X_{{\bf k}'_\|,{\bf l}_\|}
      c^{0,{\bf l}_\|,{\bf k}_\|}_{\rm plasma}
\label{eq:c_plasma}.
\end{eqnarray}
Here, the main-sum approximation has been made and the 
spontaneous recombination term is omitted since it mostly influences
the exciton population. We have also presented
the three-particle scattering terms induced by the
Coulomb scattering via $\gamma^{X}_{{\bf k}_\|,{\bf k}'_\|}$. 

The incoherent particle states are characterized by the respective single-particle
carrier distributions $f^{e/h}$ and the exciton populations $\Delta{N}_{\lambda}$.
Hence, under the assumed quasi-stationary conditions 
the PL can be solved for a given configuration by specifying
$f^{e/h}$ and $\Delta{N}_{\lambda}$.
For low densities, and if we assume that the scattering can be approximated by a simple
dephasing constant $\gamma$, Eq.~(\ref{eq:c_plasma}) reduces to
\begin{eqnarray}
\Delta N_{\lambda,\nu}^{\rm plasma}({\bf q}_\|) 
&=& 
\frac{E_\nu - E_\lambda}{E_\nu - E_\lambda -i\gamma}
	\sum_{{\bf k}_\|} 
	\phiX^L_\lambda({\bf k}_\|) 
	f^e_{{\bf k}_\|+{\bf q}_e} f^h_{{\bf k}_\|-{\bf q}_h}
	\phiX^L_\nu({\bf k}_\|)
\label{eq:XX_steady_state}
\end{eqnarray}
showing that this correlated plasma contribution does not contain densities since
the diagonal terms $\nu = \lambda$ vanish.

\subsection{Analytical approximations}
\label{sec:Elliott}

To obtain an analytical understanding of the fully numerical 
solutions of the complete equations, it is beneficial to introduce
a generalized exciton basis in order to diagonalize the homogeneous parts of
Eqs.~(\ref{eq:SBE-Pol2}), (\ref{eq:SLE_BY_steady}) and~(\ref{eq:c_plasma}). 
Since these equations include also carrier-carrier scattering, energy renormalizations 
and phase-space filling factors, a suitable exciton basis clearly must go beyond the usual
low-density Wannier wavefunctions.

The basis set for solving Eqs.~(\ref{eq:SBE-Pol2}) and~(\ref{eq:SLE_BY_steady}) is 
obtained from the solution of the generalized Wannier equation,
\be
\sum_{{\bf k}'_\|} \matrix{M}_{\vec{k}_\|,\vec{k}'_\|}(\omega) \phi^{R}_{\nu}({\bf k'}_\|,\omega)
= 
(E_{\nu}(\omega) - i \Gamma_{\nu}(\omega)) \phi^{R}_{\nu}({\bf k}_\|,\omega)
\label{eq:genWannier}
\ee
with the matrix $\matrix{M}$ defined as
\bea
M_{\vec{k}_\|,\vec{k}'_\|}(\omega) 
& = &  [ \tilde{\epsilon}_{{\bf k}_\|} - i \gamma^{\mrm{D}}_{\vec{k}_\|}(\omega)] \delta_{{\bf k}'_\|,{\bf k}_\|}
- \left[1-f^e_{{\bf k}_\|}-f^h_{{\bf k}_\|}\right] 
 V_{{\bf k}'_\|-{\bf k}_\|}
+ i \gamma^{\mrm{OD}}_{\vec{k}_\|,\vec{k}'_\|}(\omega)
 \label{eq:def_M},
\eea
Here, the scattering matrices $\gamma^{\mrm{D/OD}}(\omega)$ are given in App.~\ref{app:scat}. Due
to their occurence and because of
the phase-space filling factor, Eq.~(\ref{eq:genWannier}) is clearly 
non-Hermitian such that we have to distinguish between left- and right-handed basis functions. The
left-handed functions $\phi^{L}_{\nu}({\bf k}_\|,\omega)$ satisfy the equation 
\be
\sum_{{\bf k}'_\|} \phi^{L}_{\nu}({\bf k'}_\|,\omega) \matrix{M}_{\vec{k}'_\|,\vec{k}_\|}(\omega) 
= 
(E_{\nu}(\omega) - i \Gamma_{\nu}(\omega)) \phi^{L}_{\nu}({\bf k}_\|,\omega)
\label{eq:genWannier2}
\ee
with the same eigenvalues as the right-handed ones. Since the scattering for the polarization and the
photon-assisted polarizations is frequency dependent, also the resulting basis sets
$\{\phi^{L/R}_{\nu}({\bf k}_\|,\omega)\}$ and the complex eigenvalues $E_{\nu}(\omega) 
- i \Gamma_{\nu}(\omega)$ depend on $\omega$. 

Introducing this generalized exciton basis in Eqs.~(\ref{eq:SBE-Pol2}) and~(\ref{eq:absorption}),
we obtain the equivalent formulation
%
\be
\alpha(\omega) \propto
|d_{\mrm{cv}}|^2
\sum_{\nu,\mu}
\frac%
{\phi^R_{\nu}(r=0,\omega) \bigl(\phi^R_{\mu}(r=0,\omega)\bigr)^*}
{E_{\nu}(\omega) - \hbar \omega - i \Gamma_{\nu}(\omega)}
A_{\nu,\mu}(\omega),
\label{eq:chi}
\ee
%
with the driving term $A_{\nu,\mu}(\omega)$ defined as
%
\bea
A_{\nu,\mu}(\omega)
&=&
\sum_{\vec{k}_\|}
\left(\phi^L_{\nu}(\vec{k}_\|,\omega)\right)^\star (1 - f^e_{\vec{k}_\|} - f^h_{\vec{k}_\|})  \phi^L_{\mu}(\vec{k}_\|,\omega)
\nonumber\\
&=&
\langle \phi^L_{\nu}(\omega) | (1 - f^e - f^h) \phi^L_{\mu}(\omega) \rangle 
\label{eq:A_munu}
\eea
%
where we introduced the symbolic bra-ket notation for the inner product.

In the limit of constant $\gamma$, the relation
%
\be
(1 - f^e_{\vec{k}_\|} - f^h_{\vec{k}_\|})  \phi^L_{\mu}(\vec{k}_\|,\omega)
= \phi^R_{\mu}(\vec{k}_\|,\omega)
\label{eq:phiL_phiR}
\ee
%
can be proven such that $A_{\nu,\mu}(\omega)$ reduces to 
%
\be
A_{\nu,\mu}(\omega)\Bigr|_{\gamma}
=
\sum_{\vec{k}_\|}
\left(\phi^L_{\nu}(\vec{k}_\|,\omega)\right)^\star  \phi^R_{\mu}(\vec{k}_\|,\omega)
= \delta_{\mu,\nu}
\label{eq:A_munu_gamma}
\ee
%
and we recover the usual Elliott formula. However, in the present paper, we always
solve the absorption with the full source term, Eq.~(\ref{eq:A_munu}).

Due to the structural similarity between the semiconductor Bloch Eq.~(\ref{eq:SBE-Pol2})
and the luminescence Eq.~(\ref{eq:SLE_BY_steady}) and the identical form of the microscopic
scattering matrices, the photon-assisted polarizations $\Pi_{\vec{k}_\|,q}$ can be expanded
using the same exciton basis functions. Inserting the solution of Eq.~(\ref{eq:SLE_BY_steady})
into Eq.~(\ref{eq:SLE_b*b_steady}), we obtain
%
\be
I^{\mrm{PL}}(\omega) \propto
|d_{\mrm{cv}}|^2
\sum_{\nu,\mu}
\frac%
{\phi^R_{\nu}(r=0,\omega) \bigl(\phi^R_{\mu}(r=0,\omega)\bigr)^*}
{E_{\nu}(\omega) - \hbar \omega - i \Gamma_{\nu}(\omega)}
S_{\nu,\mu}(\omega),
\label{eq:PL}
\ee
%
which is structurally identical to Eq.~(\ref{eq:chi}) and exhibits the same resonances
and the same frequency dependent broadenings. The only difference lies in the 
PL source term
%
\be
S_{\nu,\mu}(\omega)
=
\sum_{\vec{k}_\|}
\left(\phi^L_{\nu}(\vec{k}_\|,\omega)\right)^\star 
\biggl(
  f^e_{{\bf k}_\|} f^h_{{\bf k}_\|}
        +       
        \sum_{{\bf k}'_\|}
        c_X^{0,{\bf k}_\|,{\bf k}'_\|}
\biggr)
\phi^L_{\mu}(\vec{k}_\|,\omega).
\label{eq:def_S_numu}
\ee
%
As we noted already in the discussion after Eq.~(\ref{eq:SLE_BY_steady}), the PL
source consists of two different contributions. The first one defines the Hartree-Fock 
exciton correlation
%
\be
S^{\mrm{HF}}_{\nu,\mu}(\omega)
=
\ev{X^{\dagger}_{\nu}  X_{\mu}}(\omega)\Bigr|_{HF}
=
\sum_{\vec{k}_\|}
\left(\phi^L_{\nu}(\vec{k}_\|,\omega)\right)^\star 
  f^e_{{\bf k}_\|} f^h_{{\bf k}_\|}
\phi^L_{\mu}(\vec{k}_\|,\omega).
\label{eq:def_X+X_HF}
\ee
%
It contains both diagonal and non-diagonal contributions and is by itself not a consistent
approximation for calculating PL spectra. Nevertheless, we will see later that it
still gives a qualitative picture around the $1s$ resonance.

The second contribution looks exactly like the exciton transformation defined
in Eq.~(\ref{eq:XX_transform}). However, before proceeding further, we first look 
at the solution of the last remaining Eq.~(\ref{eq:c_plasma}), which determines
the electron-hole correlations entering in the source of Eq.~(\ref{eq:def_S_numu}).
This equation can also be simplified by introducing a suitable electron-hole pair
basis, but the scattering for the $c_X$-equation is different
from the scattering entering in the previous two equations. In particular, the scattering
matrices are not frequency dependent since we assume stationary excitonic correlations.
Therefore, Eq.~(\ref{eq:c_plasma}) should be diagonalized with an exciton basis obtained
by using the correct excitonic scattering matrices in the matrix in Eqs.~(\ref{eq:def_M})
and~(\ref{eq:genWannier}). We denote this second exciton basis set by 
$\phiX^{L/R}_{\mu}(\vec{k}_\|)$. Expressing the excitonic correlations $c_X$ in the
PL source, Eq.~(\ref{eq:def_S_numu}) with this excitonic basis, we can rewrite
the source term as
%
\be
\Delta{S}_{\nu,\mu}(\omega)
=
\sum_{\nu',\mu'}
\langle 
\phi^L_{\nu}(\omega) | \Phi^R_{\nu'}
\rangle 
\Delta\ev{X^{\dagger}_{\mu'} X_{\nu'}}
\langle 
\Phi^R_{\mu'} | \phi^L_{\mu}(\omega)
\rangle 
\label{eq:DeltaS_numu}
\ee
%
where the brackets denote either the expectation value or an inner product. Thus,
the total luminescence source is given by the sum of Eqs.~(\ref{eq:def_X+X_HF})
and~(\ref{eq:DeltaS_numu}). In the spirit of Eq.~(\ref{eq:X_vs_plasma}), the
correlated source $\Delta{S}_{\nu,\mu}(\omega)$ can describe a correlated
plasma or true exciton populations. In general, we refer to plasma PL, if we 
include only the correlated plasma from Eq.~(\ref{eq:c_plasma}).to the
PL source $\Delta{S}_{\nu,\mu}(\omega)$. True excitonic PL is obtained
after we add additional excitons $\Delta{N}_{\nu} = \Delta\ev{X^{\dagger}_{\nu} X_{\nu}}$
to the source term.

In all our computations, we include the Hartree-Fock source as well as the
plasma correlations including the correct microscopic scattering,
unless stated otherwise. However, in order to gain some
more physical insight to the meaning of Eq.~(\ref{eq:PL}) {\bf and only for that
purpose, not for the numerical evaluations}, one may simplify the
microscopic scattering by replacing it with a phenomenological decay constant $\gamma$.
As a result, the exciton basis sets $\phi_\nu$ and $\phiX_\nu$ become identical and
basis as well as eigenvalues become independent of $\omega$.
Furthermore, we may apply the result from Eq.~(\ref{eq:XX_steady_state}) for the
correlated electron-hole plasma. By inserting it into Eqs.~(\ref{eq:PL})--(\ref{eq:def_S_numu})
and using Eq.~(\ref{eq:def_X+X_HF}), we obtain a simple Elliott-like formula for
luminescence,
%
\be
I^{\mrm{PL}}_{\mrm{Elliott}}(\omega)
 \propto
|d_{\mrm{cv}}|^2
\sum_{\nu}
\frac%
{|\phi^R_{\nu}(r=0)|^2 S^{\mrm{Elliott}}_{\nu} }
{E_{\nu}(\omega) - \hbar \omega - i \gamma}
\label{eq:PL_Elliott}
\ee
%
with the source term given by
%
\be
S^{\mrm{Elliott}}_{\nu}
=
\biggl(
\sum_{\vec{k}_{\|}}
|\phi^{L}_{\nu}(\vec{k}_{\|})|^2 f^e_{{\bf k}_\|} f^h_{{\bf k}_\|}
\biggr)
+
\Delta\ev{X^\dagger_{\nu,0} X_{\nu,0}}
=
N_{\nu}^{\mrm{HF}}(q=0)
+ 
\Delta{N}_{\nu}(q=0).
\label{eq:S_munu_Elliott}
\ee
%
We now see very clearly that excitonic PL can result either
from excitons or directly from 
electron and hole densities. The form 
of Eq. (\ref{eq:S_munu_Elliott}) is very useful when one envisions 
the differences of plasma and exciton PL. For example,
Ref.~\onlinecite{Kira:99} shows that plasma PL from Fermi-Dirac 
distributions produces an exponentially decaying tail for energies
above the fundamental gap. This feature survives also when the 
scattering is fully included such that one can deduce rather accurately
the temperature of electron and hole distributions directly
from the PL spectra.

\section{Numerical results}
\label{sec:results}
%
The full microscopic form of the semiconductor absorption and luminescence equations
shows that excitonic resonances can appear under quite different 
conditions. Clearly, Coulombic scattering can strongly broaden these resonances. We
also notice that the scattering terms contain highly nonlinear
contributions of the carrier densities. We therefore start by investigating these 
excitation-induced
dephasing effects\cite{Jahnke:96,Jahnke:97} on different excitonic resonances.

\subsection{Excitation-induced dephasing}
\label{sec:EID}
%
%
\begin{figure}[ht]
\resizebox{0.4\textwidth}{!}{%
\includegraphics{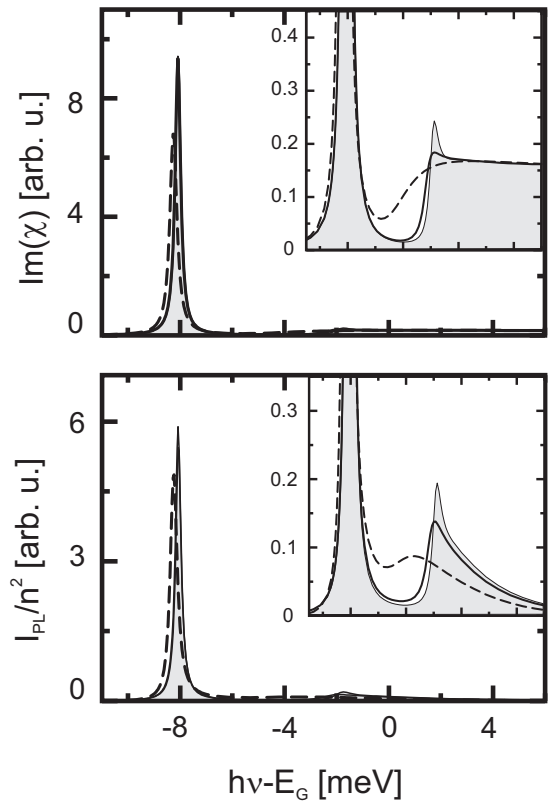}}
\caption{Absorption (top) and PL spectra (bottom) for a carrier temperature of 
$T=40$\,K for three different carrier densities of $n=5 \times 10^8$\,cm$^{-2}$ (shaded area), 
$n= 1 \times 10^9$\,cm$^{-2}$ (solid line), and $n=6 \times 10^9$\,cm$^{-2}$ (dashed line).
The insets show the magnification of the region around the $2s$ resonance.}
\label{fig:AbsAndPL}
\end{figure}
%
We first illustrate the influence of the microscopic scattering on the higher bound
states by evaluating the fully microscopic absorption and PL spectra for various 
carrier densities. More specifically, we assume Fermi-Dirac distributions at 40\,K
for electrons and holes and take three very low values for the densities which are usually
all considered to be in the ``linear regime''. In addition to the microscopic
Coulomb scattering, we have included a small background $\gamma=0.125$\, meV which
results from the effects of acoustic phonon scattering and the purely radiative decay. 
Figure~\ref{fig:AbsAndPL} presents absorption and PL spectra side by side showing
that the $1s$-exciton resonance lies 8\,meV below the band
gap. For the densities used here, it is hardly changed and its width is basically
determined by the background decay constant. But while the $1s$-exciton resonance 
is almost unchanged even for the largest density used here, the magnifications in the
insets clearly demonstrate a very strong broadening of the 2s-exciton resonance. 
The absorption spectra show that the $2s$ resonance merges
with the continuum already at around $10^9$\,cm$^{-2}$, a density which
is usually considered very low in typical GaAs-type QW systems. The same trend is
visible in the PL spectra where it is almost impossible to detect a clear $2s$
resonance, instead it merely marks the onset of the exponential decay into the 
continuum. This fact explains why the $2s$ resonance is often not cleary observable
in experiments.

The relation between excitonic resonances and nonlinearities can be investigated
most easily if we define the separate contributions of Eqs.~(\ref{eq:chi})
and~(\ref{eq:PL}) via
%
\bea
\alpha(\omega) &=&
\sum_{\nu}
\alpha_{\nu}(\omega)
\\
I^{\mrm{PL}}(\omega) & = &
\sum_{\nu}
I^{\mrm{PL}}_{\nu}(\omega)(\omega)
\eea
%
with 
%
\bea
\alpha_{\nu}(\omega) &\propto&
|d_{\mrm{cv}}|^2
\sum_{\mu}
\frac%
{\phi^R_{\nu}(r=0,\omega) \bigl(\phi^R_{\mu}(r=0,\omega)\bigr)^*}
{E_{\nu}(\omega) - \hbar \omega - i \Gamma_{\nu}(\omega)}
A_{\nu,\mu}(\omega),
\label{eq:alpha_nu}
\\
I^{\mrm{PL}}_{\nu}(\omega) &\propto&
|d_{\mrm{cv}}|^2
\sum_{\mu}
\frac%
{\phi^R_{\nu}(r=0,\omega) \bigl(\phi^R_{\mu}(r=0,\omega)\bigr)^*}
{E_{\nu}(\omega) - \hbar \omega - i \Gamma_{\nu}(\omega)}
S_{\nu,\mu}(\omega).
\label{eq:PL_nu}
\eea
Here, the individual contributions of the different exciton
states $\nu$ appear separately. However, since the eigenenergies 
and broadenings are frequency dependent, this simple form is in fact somewhat
misleading. Due to the frequency dependence of the complex excitonic eigenvalues
the generalized Lorentzian function can in reality deviate quite strongly
from a simple Lorentzian --- even more so since in general the oscillator
strength in the numerator can be complex and thus mix real and imaginary parts
of the energy denominator.

\begin{figure}[ht]
\resizebox{0.4\textwidth}{!}{%
\includegraphics{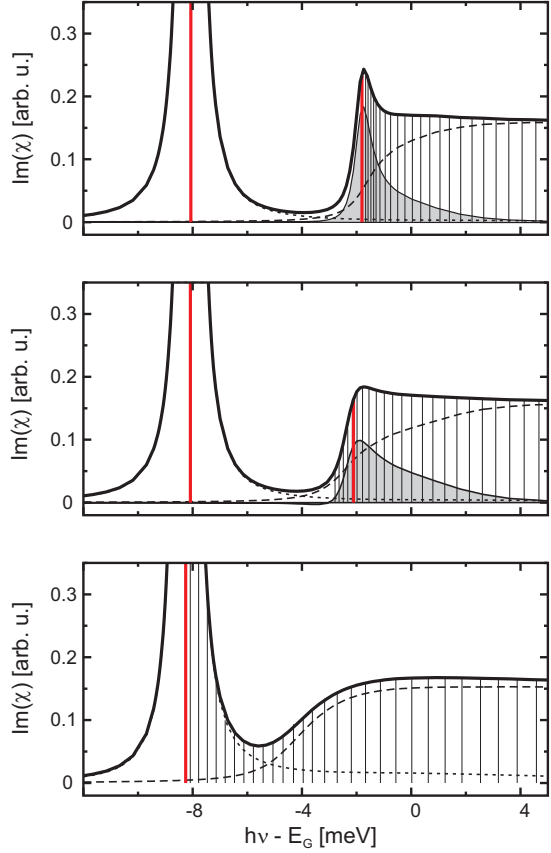}}
\caption{Absorption spectra for a carrier temperature of $T=40$\,K for three different
densities of $n=5 \times 10^8$\,cm$^{-2}$ (top), $n= 1 \times 10^9$\,cm$^{-2}$ (middle),
and $n=6 \times 10^9$\,cm$^{-2}$ (bottom). In addition to the full spectrum (solid line),
the separate contribution for the $1s$ resonance (dotted line), the $2s$ resonance (shaded area),
and the rest (dashed line) is given. The vertical lines display the real parts of the
excitonic eigenenergies of the generalized Wannier Eq.~(\ref{eq:genWannier}).}
\label{fig:AbsDetail}
\end{figure}
In Fig.~\ref{fig:AbsDetail}, we show the same three absorption spectra as
in Fig.~\ref{fig:AbsAndPL} as thick solid lines. But now, we also plot
$\alpha_{1s}$, $\alpha_{2s}$ and the rest separately. Since the excitonic
states have a complicated $\omega$ dependence due to the scattering matrices,
we identified the $1s$ and $2s$ resonances by choosing at each $\omega$
those two excitonic states which have the highest oscillator strength. Especially
for higher densities, we find an entire continuum of excitonic states; some
of them have a relatively small oscillator strength. But a clear 
distinction of the different states, as in the zero-density analytical Wannier formula, is no
longer possible.

Even for the lowest
density, the $2s$ resonance is strongly non-Lorentzian and extends well
into the continuum. Reversely, even the higher lying continuum states
contribute significantly
(almost a quarter) to the observed $2s$ peak. For the intermediate
density, this effect becomes even more pronounced and for the highest density,
it is no longer possible to determine a clear resonance. This becomes 
understandable if we look at the vertical lines 
in the figure which indicate the positions
of the real parts of the eigenvalues. For low densities these
real parts agree with the expected energies of the $1s$, $2s$ etc. states. However, we 
see that for elevated densities the microscopic scattering leads to resonances 
even between the $1s$ and $2s$ energies. For the highest density, these
resonances can extend down to the $1s$ peak itself.

%
\begin{figure}[ht]
\resizebox{0.4\textwidth}{!}{%
\includegraphics{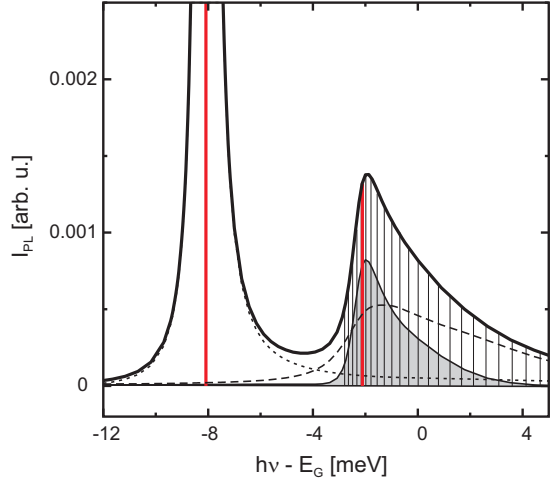}}
\caption{PL spectrum and the different contributions for the same parameters 
as the middle frame of Fig.~\ref{fig:AbsDetail}.}
\label{fig:PLDetail}
\end{figure}
Exactly the same analysis as before is performed for our PL spectra. For the
intermediate density, the result is displayed in Fig.~\ref{fig:PLDetail}. All
features discussed in Fig.~\ref{fig:AbsDetail} are clearly visible also here.
The excitonic positions are the same as in the absorption analysis, since the exciton basis for
semiconductor Bloch and luminescence equations are identical. As we have already
mentioned in the discussion of Fig.~\ref{fig:AbsAndPL} a clear $2s$ resonance
cannot be distinguished. It is composed of a whole cluster of resonances and
has to be analyzed with a full microscopic theory. A zero-density Wannier
basis is clearly insufficient to understand these intricate effects.

%
\begin{figure}[ht]
\resizebox{0.4\textwidth}{!}{%
\includegraphics{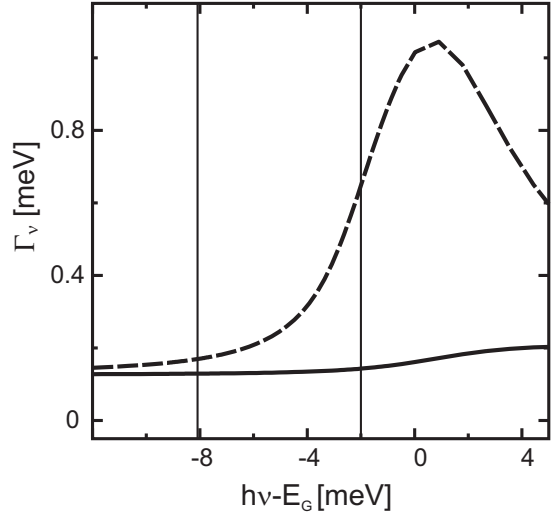}}
\caption{Imaginary part of excitonic eigenenergy of the generalized Wannier 
Eq.~(\ref{eq:genWannier}) for the two bound states with the hightest (solid line) and 
the second-highest (dashed line) oscillator strength for a carrier temperature of 
$T=40$\,K and a carrier density of $n= 1 \times 10^9$\,cm$^{-2}$. The vertical lines
indicate the respective real parts.}
\label{fig:Gamma}
\end{figure}
Figure~\ref{fig:Gamma} presents an important part of the information necessary to understand
the strong non-Lorentzian lineshape features. Here, we show the frequency dependence of
the broadening of the two bound states with the highest oscillator strength. 
For simplicity, we continue to call the second resonance a $2s$ resonance in 
analogy to the usual Wannier nomenclature. Our observation from the spectra
are confirmed here; while the $1s$ resonance is almost independent of frequency
and dominated by the background decay constant $\gamma_{\mrm{bg}}=0.125$\,meV, the
broadening of the second resonance is strongly frequency dependent. At the frequency
value where the maxima of the corresponding resonances are observed (indicated by
the vertical lines) the broadening of the second
is already about five times as big as that of the first
resonance. The fact that this broadening decays more quickly towards lower frequencies than 
towards higher frequencies explains the non-Lorentzian feature and the larger
extension of the $2s$ resonance towards the continuum states. At the same time, 
the small $1s$ broadening is consistent with the observed narrow Lorentzian
line from Figs.~\ref{fig:AbsAndPL}--\ref{fig:PLDetail}.

This explanation covers only part of the non-Lorentzian lineshape. By looking
at Eqs.~(\ref{eq:chi}) and~(\ref{eq:PL}) we note that fundamentally the influence
of scattering can be even more complicated. Since the presence of scattering
generally leads to complex wave functions and correlations $A_{\nu,\mu}$ and
$S_{\nu,\mu}$, even the oscillator strength can be complex and real and imaginary
parts, i.e.\ broadened $\delta$-function and principal value contributions in the complex
Lorentzian functions can be mixed. 

Before turning our attention to more details of the PL spectra, we summarize the
important finding of this section. We have noted that both in absorption and in 
PL  the so-called ``$2s$ resonance'' actually consists of a whole cluster of energetically
close resonances. Except for ultra-low densities the usual nomenclature
fails and the $2s$ resonance is not necessarily the second lowest bound state.
Microscopic Coulomb scattering leads to the presence of a continuum of states,
sometimes even relatively close to the $1s$-exciton resonance.

\subsection{Plasma luminescence spectra}
\label{sec:lumi}
%
Since we have seen in the previous section that the analysis of the
$2s$ resonance is a rather delicate
issue, we start our detailed investigation of PL spectra by comparing different
approximations and their influence around the $1s$ resonance. This resonance is modified
much less by Coulomb scattering than the $2s$ resonance
such that we can anticipate it to be more robust 
under different approximations applied. 

In the present section, all PL spectra are computed from Eq.~(\ref{eq:PL}) with the
Hartree-Fock source, Eq.~(\ref{eq:def_X+X_HF}), and the correlated plasma contribution
obtained from Eq.~(\ref{eq:c_plasma}). We do not add additional diagonal exciton
populations until the next section where the ratio between excitonic and plasma
luminescence is investigated. For comparison, we also present solutions of the
simplified Elliott formula, Eq.~(\ref{eq:PL_Elliott}), where the decay constant
is chosen such as to match the width of the $1s$ resonance of the full computation.
Within the Elliott formula approach, there are three different possibilities
to deal with the four-point $c_X$ correlations. The best and most consistent way
is the full solution of Eq.~(\ref{eq:c_plasma}) with the scattering matrices
replaced by a constant $\gamma$. The other two possibilities truncate already 
part of the four-point terms and can thus lead to inconsistencies at higher bound
or continuum states. We include them here for comparison only and in order to 
relate previous results to the full theory. In the simplest case, we can
neglect the $c_X$ correlations completely. Then we obtain spectra in the same approximation
as in our first publications\cite{Kira:97}, referred to in the present paper as
``Hartree-Fock'' approximation. We can improve this approximation by solving the
four-point correlations in a scattering approach as has been discussed in Sec.~4.4
of Ref.~\onlinecite{Kira:99}. In that case, the four-point correlations are 
obtained as 
%
\be
c^{0,{\bf k}'_\|,{\bf k}_\|}_{\rm X}\Bigr|_{\mrm{scat}}
=
\frac{%
-S^{0,{\bf k}'_\|,{\bf k}_\|}}
{\epsilon^{{\bf q}_\|,{\bf k}'_\|,{\bf k}_\|} - i \gamma }
\label{eq:cX_scat}
\ee
%
with the definitions from Eqs.~(\ref{eq:E_exciton}) and~(\ref{eq:S_X}). As can
be seen from the energy denominator, Coulomb resonances are not included
in the description of the electron-hole correlations in that approximation.
Hence, we have four different ways to calculate the $1s$ PL: the fully
consistent plasma result from Eq.~(\ref{eq:c_plasma}) with microscopic scattering
matrices, the same with a constant $\gamma$ approximation (Elliott), and then two 
approximations where either Eq.~(\ref{eq:c_plasma}) is truncated and solved 
without the Coulomb sums (scattering approximation) or where the four-point
correlations are neglected completely (Hartree-Fock).

%
\begin{figure}[ht]
\resizebox{0.4\textwidth}{!}{%
\includegraphics{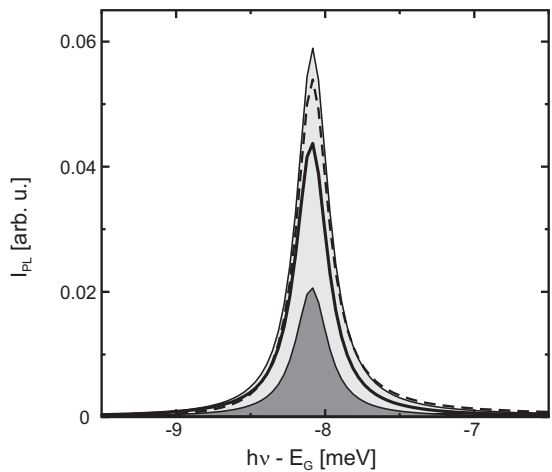}}
\caption{PL spectrum in the spectral vicinity of the $1s$ resonance. The full result
from Eq.~(\ref{eq:PL}) (shaded area) is compared to three different approximations
of the Elliott formula with constant dephasing, Eq.~(\ref{eq:PL_Elliott}). Here, the correlated
PL source term is treated fully (solid line), in scattering approximation (dashed line), or
set to zero (Hartree-Fock approximation, dark-shaded area). The constant-gamma results have 
been shifted by 40\,$\mu$eV in order to collocate the peaks at the same frequency.}
\label{fig:PL1s}
\end{figure}
\begin{figure}[ht]
\resizebox{0.4\textwidth}{!}{%
\includegraphics{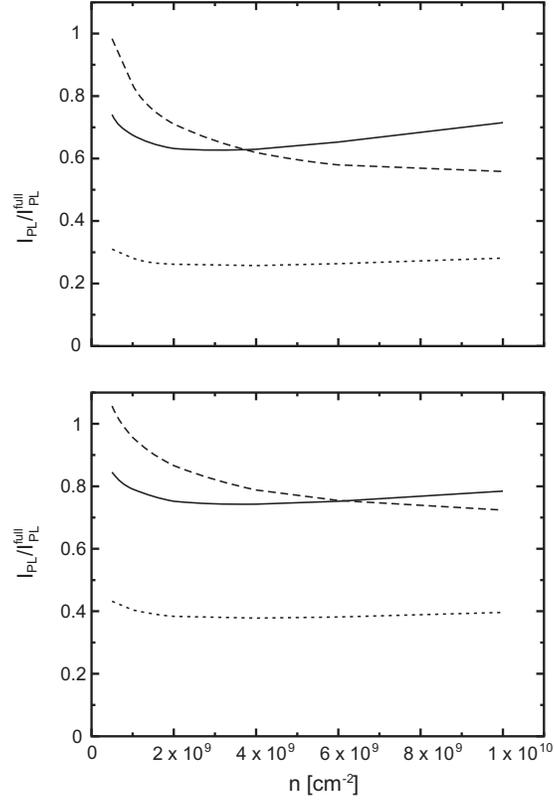}}
\caption{Relative peak height of Elliott formula (solid line), second-Born source (dashed line),
and Hartree-Fock source term (dotted line) relative to the peak height of the full plasma
result with microscopic scattering. Results for carrier temperature of $T=20$\,K (top)
and  $T=60$\,K (bottom) are shown.}
\label{fig:Ratio1s}
\end{figure}

Figure~\ref{fig:PL1s} presents the PL spectra around the $1s$ resonance for 
these four different ways of computing the excitonic correlations. All approximations
exhibit a clear peak at the $1s$ resonance. A small shift of 40\,$\mu$eV  for the
three constant-gamma calculations compensates small differences in the energy
renormalizations with and without microscopic scattering. While the Hartree-Fock
approximation gives only about one third of the full plasma PL, the other
two approximations are within 30\% of the full plasma value. Since we adjusted
the decay constant, we can directly compare the peak heights in order to compare
the quality of the different approximations. This is done in detail in Fig.\ref{fig:Ratio1s}
for the two carrier temperatures of $T=20$\,K and $T=60$\,K. For both temperatures,
the Eliott formula approaches the full plasma PL for very low densities, deviates 
most strongly at some intermediate densities, before getting better again in the limit 
of higher densities. The maximum deviation is temperature dependent and is
less than 40\% for 20\,K and even smaller around  25\% for 60\,K. The Hartree-Fock
approximation shows a similar functional form, but is in general too small by 
a factor of around three to four, with the precise value also depending on carrier
density and temperature. The scattering approximation decays monotonously from
low to high densities. For very low densities, it can even overestimate the correct
value of the plasma luminescence as can be anticipated by extrapolating the curve
to zero density. For high densities, it is too small by less than a factor of two.
%

\subsection{Excitonic plasma luminescence}
\label{sec:nonTDplasmaPL}

%
\begin{figure}[ht]
\resizebox{0.4\textwidth}{!}{%
\includegraphics{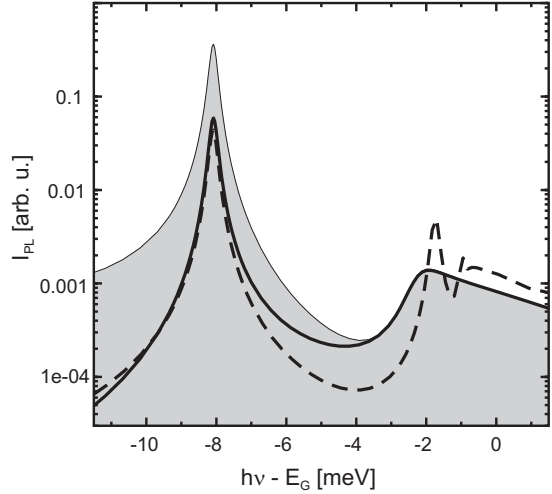}}
\caption{PL spectra for the same parameters as in Fig.~\ref{fig:PL1s} in semilogarithmic
plot. The full plasma PL (solid line) and the Elliott-formula result are compared to
the thermodynamic limit (shaded area).}
\label{fig:PLlog}
\end{figure}
The results in Sec.~\ref{sec:lumi} show that the intensity of the 
$1s$-plasma PL can be predicted rather accurately using different levels of
approximations. We next quantify the characteristics of the
$1s$ PL with respect to other resonances in the spectrum.
Especially, we may use the thermodynamic relation (\ref{eq:KMS})
between $\alpha(\omega)$ and ${\rm PL}(\omega)$ to define how
nonthermal the excitonic plasma PL is.
The computed full (solid line), Elliot (dashed line), and thermodynamic 
(shaded area) PL are presented in Fig.~\ref{fig:PLlog}. Here, the carrier density
has been taken as $10^9 {\rm cm}^{-2}$ and the lattice temperature is 40~K.
The shown thermodynamic PL
is simply obtained by computing the microscopic absorption from
Eq.~(\ref{eq:chi}). Then, $\alpha(\omega)$ is multiplied by
a thermal factor according to Eq.~(\ref{eq:KMS}). The temperature 
is chosen such that the high-energy PL tails of the full and the 
thermodynamic computations agree.

We observe now that event though the full and the Elliot PL agree at the
$1s$ resonance, the Elliot PL overestimates the high-frequency part of the
spectrum since microscopic Coulomb scattering is not included.
This is a clear indication that one needs to include the microscopic
scattering to understand the quantitative features in the full 
spectral range of the excitonic PL. A comparison of the full and the
thermodynamic PL indicates that the excitonic plasma PL
is strongly nonthermal since it produces a much weaker $1s$ resonance than
thermodynamic arguments predict. We will see, later on, that one actually
needs true $1s$-exciton populations to enhance the $1s$ PL toward
its thermodynamic value. The inherent nonequilibrium feature of the
excitonic plasma luminescence has a simple explanation. It is intuitively
clear that $1s$ emission from plasma or correlated electron-hole
plasma is more complicated than from exciton populations. 
This complication arises from the fact that an electron-hole plasma can emit a photon at
excitonic frequencies, which are much below its average energy per particle,
only by heating the remaining plasma.
Due to this additional difficulty, the excitonic PL resulting 
from a pure plasma is generally 
weaker than that for the pure thermodynamic limit containing also exciton
populations.

%
\begin{figure}[ht]
\resizebox{0.4\textwidth}{!}{%
\includegraphics{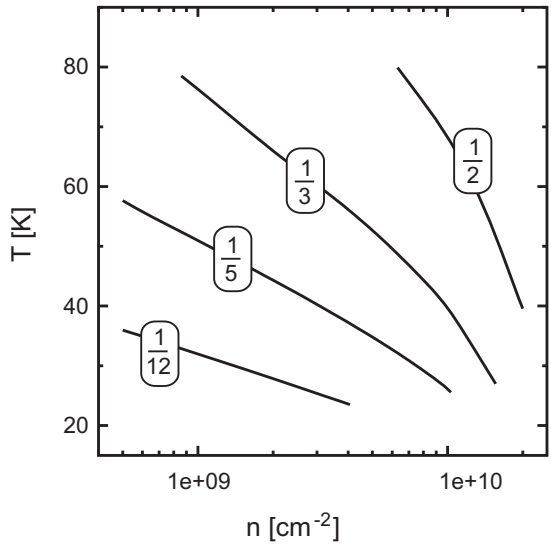}}
\caption{``Phase-space'' plot for $\beta$-factor as function of carrier density and
temperature. The temperature has been determined from the continuum tail of the emission.}
\label{fig:beta}
\end{figure}
To quantify the level of nonthermal character in more detail, we
define a $\beta$ suppression factor,
\begin{equation}
\beta \equiv \frac{{\rm PL}_{1s}|_{\rm full}}{{\rm PL}_{1s}|_{\rm TD}},
\label{eq:beta_fact},
\end{equation}
first introduced by Schnabel {\it et al.}\cite{Schnabel:92}.
Essentially, the $\beta$ factor tells us how strongly the actual $1s$ PL,
${\rm PL}_{1s}|_{\rm full}$, is suppressed with respect to
its thermodynamic value ${\rm PL}_{1s}|_{\rm TD}$.
Figure \ref{fig:beta} presents a contour plot of $\beta$ as function
of carrier density and temperature when only the fully mictroscopically
included plasma luminescence is included. As a general tendency, the plasma
PL becomes strongly nonthermal for low densities and temperatures. This
is the regime which is favourable to sustain truly bound
excitons. As temperature and/or density are increased, the excitonic plasma
PL approaches the thermodynamic limit.

As a general tendency, the $\beta$ analysis has a rather weak sensitivity
to the scattering model used or to low levels of disorder.\cite{Bozsoki:06}
Thus, the $\beta$ analysis is very much suitable to analyze the
plasma nature of experimentally detected PL. In other words,
the presented $\beta$-phase diagram can be applied to diagnose 
the degree to which the measured PL in   
low-disorder GaAs-type samples is caused by plasma emission. 
On the basis of a quantitative analysis of simultaneously measured PL
and absorption, Refs.~\cite{Chatterjee:04,Hoyer:05} have shown that
the experimental results for 
temperates above 30~K and densities above $5\times 10^{9} {\rm cm}^{-2}$
can be attributed to pure plasma emission.

\subsection{Exciton-population effects}
\label{sec:popPL}

Under suitable conditions a system of incoherent electron-hole 
quasi-particle excitations consists a mixture of correlated 
electron-hole plasma and exciton populations. 
Clearly, exciton populations may appear in all exciton states
including the ionization continuum. In particular, the full thermodynamic equilibrium
between plasma and excitons leads to a many-body 
problem \cite{Portnoi:99,Siggelkow:04}
that produces highly nontrivial many-body states. However, several recent
experiments \cite{Schnabel:92,Chatterjee:04,Hoyer:05} demonstrate
that excitonic PL is almost always far from the purely thermodynamic limit. Thus, 
it seems 
realistic to assume that the correlated electron-hole plasma coexists together
with a fraction of excitons that is determined by the exciton
formation kinetics and typically 
much below its thermodynamic value.
Since the $1s$-exciton populations dominate at low temperatures, we include
only them in our analysis of the exciton population contribution to the PL.

As discussed in Sec.~\ref{sec:nonTDplasmaPL}, the PL at the
$1s$ resonance can be strongly
enhanced by the presence of true exciton populations. This can be seen
directly from the luminescence Elliot formula (\ref{eq:PL_Elliott}).
More precisely, a $1s$-exciton population basically adds luminescence
at the $1s$ frequency whose magnitude increases
as a linear function of the exciton occupation
$\Delta N_{1s}$. In order to investigate this effect, we add
$1s$ exciton populations to the fully microscopically computed
correlated plasma luminescence. We assume that the added $1s$
excitons have a thermal momentum distribution such that
$\Delta N_{1s}(0)$, i.e. the 'bright part' of the exciton distribution that
contributes to the luminescence in normal direction, 
is known. In practice, the population $\Delta N_{1s}(0)$ is added
to the source (\ref{eq:X_vs_plasma}) such that the fully microscopic luminescence
is solved from Eqs.~(\ref{eq:PL}) for the coexisting exciton and plasma
populations.

%
\begin{figure}[ht]
\resizebox{0.4\textwidth}{!}{%
\includegraphics{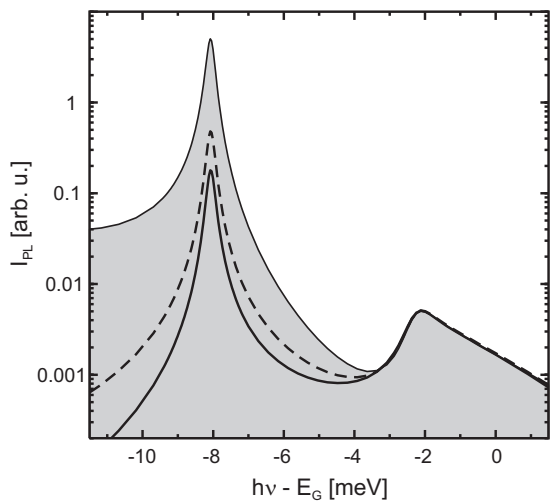}}
\caption{PL spectra for a carrier density of $n= 1 \times 10^9$\,cm$^{-2}$ and
a nominal carrier temperature of $T=20$\,K. The full plasma result (solid line) is compared
to the thermodynamic curve (shaded area). Adding 1\% excitons at a temperature of $T_X = 10$\,K
(dashed line) drastically increases the $1s$ PL.}
\label{fig:PLplusX}
\end{figure}
To illustrate the basic effect of exciton populations, we assume a low 
carrier density of $10^9 {\rm cm}^{-2}$ and 20~K carrier temperature.
Since the excitons can often reach a lower temperature than the carrier 
distributions, we choose a 10~K temperature for the excitons.
Figure \ref{fig:PLplusX} shows PL without excitons (solid line),
1\% excitons (dashed line), and the thermodynamic limit (shaded area).
We notice that in
all cases, the PL spectra are qualitatively similar, especially, the PL
above the second resonance is virtually the same. Hence, the
main feature of $1s$ populations is that they increase the $1s$ PL toward the
thermodynamic limit. This confirms our previous assumption that
exciton populations must be included to produce a thermodynamic PL.

%
\begin{figure}[ht]
\resizebox{0.4\textwidth}{!}{%
\includegraphics{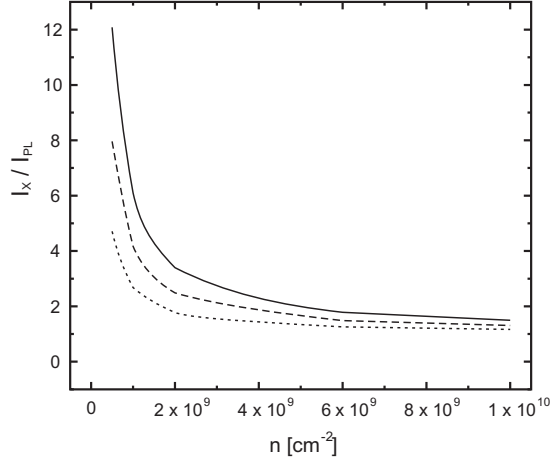}}
\caption{Ratio between $1s$ PL with and without an added amount of 1\% excitons.
For an excitonic temperature of $T_X = 10$\,K, the carrier temperature is $T=40$\,K
(solid line), $T=30$\,K (dashed line), and  $T=20$\,K (dotted line).}
\label{fig:addX_Ratio}
\end{figure}
We investigate next how much the $1s$ PL is changed by 
a 1\% fraction of $1s$ excitons as function of total carrier density.
As in Fig.~\ref{fig:PLplusX}, we assume a 10~K exciton temperature. 
For each density, we define the ratio of the $1s$ peak heights,
\begin{equation}
R_X \equiv \frac{{\rm PL}_{1s}|_{\rm X}}{{\rm PL}_{1s}|_{\rm full}},
\label{eq:R_X},
\end{equation}
between the case with excitons, ${\rm PL}_{1s}|_{\rm X}$, and
without, ${\rm PL}_{1s}|_{\rm full}$, producing the
microscopic plasma PL. The results are shown in Fig.~\ref{fig:addX_Ratio}
as function of carrier density for 20~K (solid line), 30K (dashed line),
and 40~K (dotted line) carrier temperature. The exciton populations
have strongest effect at low densities while $R_X$ approaches unity
for the elevated densities. This is in agreement with our earlier 
observation that high temperature plasma PL tends to approach the
thermodynamic limit, i.e., excitons are then less important.
However, we observe a strong increase of $R_X$ for low densities
and temperatures. According to Eq.~(\ref{eq:PL_Elliott}), the 
low-density plasma PL scales with the product of the carrier occupations,
$f^e f^h$, while the exciton population PL is directly proportional to
the occupation $\Delta N_{1s}$. When the carrier temperature is
increased or the carrier density is decreased, the absolute magnitude of
the plasma PL decreases nonlinearly. If the exciton occupation has a fixed
temperature, as we have assumed in Fig.~\ref{fig:addX_Ratio}, the relative strength of
the population PL increases with respect to the plasma PL as the density is lowered 
or the carrier temperature is increased. Thus, the observed increase of 
$R_X$ follows directly from the nonlinear dependence of the plasma
PL for low densities. In particular, the exciton populations can contribute
strongly under low temperature and low density conditions as verified
in recent experiments.\cite{Chatterjee:04,Hoyer:05}

%
\begin{figure}[ht]
\resizebox{0.4\textwidth}{!}{%
\includegraphics{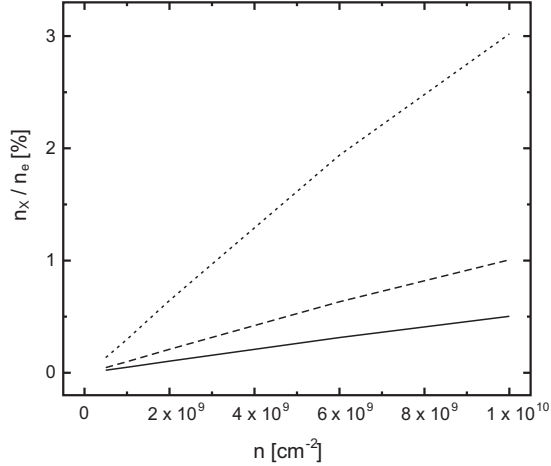}}
\caption{Necessary exciton fraction to be added in order to obtain a ratio of 3/2 between
the $1s$-resonance peaks with and without additional excitons. For an excitonic temperature 
of $T_X = 10$\,K, the carrier temperature is $T=60$\,K (solid line), $T=40$\,K (dashed line), 
and  $T=20$\,K (dotted line).}
\label{fig:FixRatio}
\end{figure}
The effect of $1s$ populations can also be investigated by evaluating
how large an exciton fraction,
\begin{equation}
x^{50\%}_{1s} \equiv
\frac{\Delta n_{1s}|_{50\%}}{n_{eh}}
\label{eq:R_50},
\end{equation}
is needed to increase the excitonic $1s$ luminescence by 50\%.
As in Fig.~\ref{fig:addX_Ratio}, we assume that the $1s$-exciton distributions follow a thermal
Bose-Einstein distribution at a temperature of 10~K. We assume carrier temperatures of
20~K (solid line), 40~K (dashed line), and 60~K (dotted line)
and define $x^{50\%}_{1s}$ as function of density. The result is displayed
in Fig.~\ref{fig:FixRatio}. We observe again that low carrier densities are 
sensitive to the presence of exciton populations while their effect becomes
less important for elevated temperatures and densities. The difference of 
low vs.~high temperature can be understood on the basis of the same nonlinearity 
arguments as in Fig.~\ref{fig:addX_Ratio}.

\section{Discussion and Conclusion}
\label{sec:discussion}

Excitonic PL resulting from an electron-hole plasma 
has been observed in a variety of 
experiments performed with GaAs-type QW systems.
For example, the ultrafast build up of the $1s$-luminescence resonance on
a subpicosecond time scale REF
is a clear indication of emission from a relaxing electron-hole plasma. 
The formation times of incoherent exciton poplulations
are much longer. This has been carefully
analyzed recently in experiments where absorption
and PL spectra have been measured 
simultaneously.\cite{Chatterjee:04,Hoyer:05} The microscopic
analysis of these experiments not only verify that
excitons are formed on a time scale of several hundred picoseconds
but they also showed that 
appreciable exciton populations are obtained only for temperatures 
below 30\,K and densities below $10^{10}$\,cm$^{-2}$. Otherwise, 
the excitonic PL could be explained by our plasma PL alone.

%
\begin{figure}[ht]
\resizebox{0.4\textwidth}{!}{%
\includegraphics{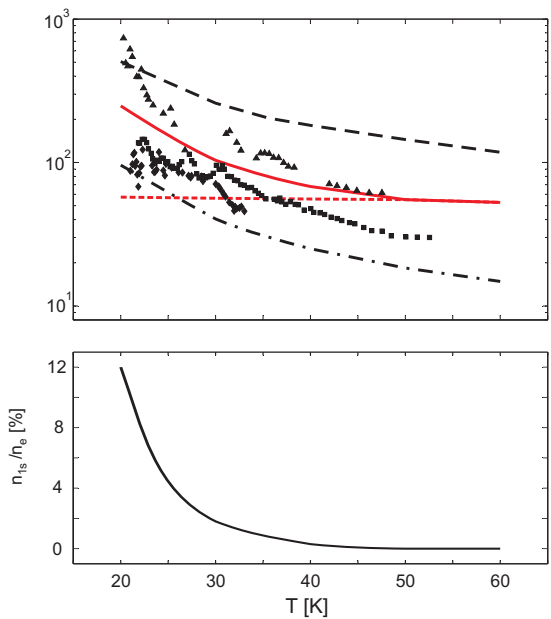}}
\caption{Top: A comparison between the data points from Fig.~5 of Phys.\ Rev.~B~{\bf 71},
195313 (2005) and the thermodynamic result as obtained from analysing either peak heights
(dashed) or areas (dot-dashed line) of $1s$ and higher resonances, respectively. Our result
without (dotted) and with (solid) added excitons obtained from the peak-height analysis
has been added. Bottom: Necessary exciton fraction for an excitonic temperature of
$T_X = 10$\,K used for the top part of the figure.}
\label{fig:Deveaud}
\end{figure}
In order to check these interpretations, we
analyze the experimental data published recently in Ref.
\cite{Szczytko:05}.
In Fig.~\ref{fig:Deveaud}, we have carefully
read in all experimental data points from Fig.~5 in Ref. \cite{Szczytko:05}
We compare these
date to our full theory and several levels of approximation.
We first compute the $1s$-to-$2s$ ratio
by integrating the thermodynamic result, Eq.~(\ref{eq:KMS}), obtained
from our absorption spectrum for a density of $10^9$\,cm$^{-2}$.
If we determine the ratio by integrating a spectral range of $8\,meV$
around each resonance as has bene done in Ref.~\onlinecite{Szczytko:05},
we obtain the dot-dashed line; if we plot the peak ratio instead, we
get the dashed line. The area ratio lies well below almost all of the
experimental data points. One can think of a large variety of possible 
explanations, which include radiative effects (the sample is measured
in reflection geometry in front of a distributed-Bragg-mirror) or 
possible disorder effects which could potentially enhance the $1s$
emission relative to the rest. In order to proceed with our analysis,
we therefore use the peak ratio as a reference, which lies clearly
above most of the experimental data points. The results of our
luminescence theory for various temperatures is plotted in 
Fig.~\ref{fig:Deveaud} before and after adding additional excitons
to our plasma PL. The necessary fraction of excitons needed
to obtain our fit is given in the bottom panel. Already the plasma
theory nicely agrees with the experimental results for elevated
temperatures above 40\,K. And for lower temperatures we find that
exciton populations do indeed contribute
significantly. Even though we do not have
access to the full experimental information, this comparison shows 
that the experimental results can be well
explained with our theory, in contrast to the claims in 
Ref.~\onlinecite{Szczytko:05}. For
a detailed one-to-one analysis, the carrier density for each data
point should be obtained in the experiment as well.

As an alternative experimental route, also THz experiments\cite{Kaindl:03} 
performed under similar conditions show unambiguously that $1s$-exciton 
populations are formed on a nanosecond time scale. 
Recently, also a combination of PL and THz absorption measurements has been 
reported.\cite{Galbraith:05} There it has been shown that a $1s$-exciton resonance
can indeed be seen in the PL even when THz absorption does not exhibit a 
$1s-2p$ resonance indicating unambiguously that $1s$-exciton populations 
are not present under these conditions.

Both the strong non-thermal nature of PL spectra and the excitonic
PL without a corresponding THz absorption signal can easily be
explained if one includes the plasma PL in the analysis.
While varying in importance, this contribution is fundamentally 
always present since it only requires that carrier distributions 
are excited. The strength of the present theory is to simultaneously treat
effects from both the electron-hole plasma and the excitons by
fully including  the Fermionic 
substructure of the excitons.
Only such a combined analysis without exclusive restriction
to one of the possible emission sources can answer the question 
which source is important under given experimental conditona.

To summarize, in the present paper 
we have discussed and evaluated our microscopic
theory for calculating PL spectra of semiconductor QW systems and
for analyzing exciton and plasma contributions to luminescence,
respectively. We have particularly focused on the effects of
microscopic Coulomb scattering and analyzed their influence
on the different exciton resonances.
While the PL Elliott formula, Eq.~(\ref{eq:PL_Elliott}),
elucidates the underlying physics, a realistic theory-experiment
comparison should always include microscopic scattering, in particular
if the detailed structure around the band-edge is of special interest.
A desirable extension of our theory would be to simultaneously
include the exciton formation dynamics and the PL in a systematic
way. Even though the theoretical framework exists, the numerical 
implementation of such a theory is very demanding and requires
a lot of additional work and improved computer resources. 

%
\begin{acknowledgments}
This work is supported by the Optodynamics Center of
the Philipps-Universit{\"a}t Marburg and by the Deutsche 
Forschungsgemeinschaft through the Quantum Optics in Semiconductors 
Research Group.
\end{acknowledgments}
\appendix
%

\section{Mathematical description of excitons}
\label{app:X_math}
%
Based on the four-point definition in Sec.~\ref{sec:def_X}, we find a natural way to identify 
incoherent excitons once we evaluate the electron-hole pair-correlation function
\begin{eqnarray}
   &&g_{\rm eh}({\bf r}_\|)
   \equiv \langle{\Psi}^\dagger_e({\bf r}_\|) 
   \Psi^\dagger_h(0) \Psi_h(0) {\Psi}_e({\bf r}_\|) \rangle
\label{eq:def_pair-corr},
\\
&&{\Psi}_e({\bf r}_\|) = \frac{1}{\sqrt{\sf S}} 
   \sum_{{\bf k}_\|}
   a_{c,{\bf k}_\|} e^{i{\bf k}_\| \cdot {\bf r}_\|},
\;\;\;\;
{\Psi}^\dagger_h({\bf r}_\|) = \frac{1}{\sqrt{\sf S}} 
   \sum_{{\bf k}_\|}
   a_{v,{\bf k}_\|} e^{i{\bf k}_\| \cdot {\bf r}_\|}
\label{eq:macr_phi},
\end{eqnarray}
where only the envelope part of the field operator
$\hat{\Psi}_\lambda({\bf r}_\|)$ is included since the
Wannier excitons depend on a length scale much longer than
the lattice periodic part of the Bloch functions.
In general, $g_{\rm eh}({\bf r}_\|)$ defines the conditional
probability to find an electron at position ${\bf r}_\|$
when the hole is positioned at the origin. Thus, properties
of $g_{\rm eh}({\bf r}_\|)$ can identify what kind of
correlations one can observe in the relative coordinate between
the electrons and holes in the many-body system.

The pair-correlation function can always be divided into single-particle
as well as to correlated two-particle contributions and 
the homogeneous incoherent quasiparticle
excitation conditions produce 
\begin{eqnarray}
   &&g_{\rm eh}({\bf r}_\|)
   \equiv  n_e n_h + \Delta g_{\rm eh}({\bf r}_\|)
\label{eq:def_pair-corr_FIN},
\\
&&\Delta g_{\rm eh}({\bf r}_\|)
\equiv 
\frac{1}{{\sf S}^2} 
\sum_{{\bf k}_\|,{\bf k}'_\|,{\bf q}_\|}
c_X^{{\bf q}_\|,{\bf k}'_\|,{\bf k}_\|}
 e^{i({\bf k}'_\| +{\bf q}_\| -  {\bf k}_\|) \cdot {\bf r}_\|}
\label{eq:pair-corr_D}.
\end{eqnarray}
We observe now that the single-particle factorization
leads to a background contribution, $ n_e n_h$,
which states that probability of finding electron and hole simultaneously
is proportional to the product of their respective densities.
This contribution does not have a genuine ${\bf r}_\|$-dependence.
As a result, the true exciton effects are observed only
in the correlated two-particle contributions 
$\Delta g_{\rm eh}({\bf r}_\|)$ which contains exclusively the
$c_X$ correlation. Thus, it is obvious now that only $c_X$ 
contributions can define and include the microscopic description
of the true excitons in the many-body system.

The correlated $\Delta g_{\rm eh}({\bf r}_\|)$ can also be expressed
in the exciton basis
\begin{eqnarray}
&&\Delta g_{\rm eh}({\bf r}_\|)
=
   \sum_{\lambda,\nu}
   \left(
	{ \frac{1}{\sf S}}
	\sum_{{\bf q}_\|}
	\Delta N_{\lambda,\nu}({\bf q}_\|) 
	\right)
	\left[
	\phi^R_\lambda({\bf r}_\|) 
	\right]^\star
	  \phi^R_\nu({\bf r}_\|)
\label{eq:g_eh_Xbas}.
\end{eqnarray}
When $\Delta g_{\rm eh}({\bf r}_\|)$ exhibits a clear single-state 
$r$-dependence proportional to $|\phi^R_{\lambda_0}({\bf r}_\|)|^2$, 
it is obvious that the $\sum_{\lambda,\nu}$ sum is dominated by the element 
$\lambda = \nu = \lambda_0$. Since this case corresponds to having
only $\lambda_0$ excitons, we may conclude that 
\begin{eqnarray}
&&\Delta n_{\lambda_0} \equiv
	{ \frac{1}{\sf S}}
	\sum_{{\bf q}_\|}
	\Delta N_{\lambda_0,\lambda_0}({\bf q}_\|) 
\label{eq:X_density}
\end{eqnarray}
defines the density of excitons in state $\lambda_0$ in the system.

\section{Microscopic scattering}
\label{app:scat}
%
In order to get a realistic density dependence for the saturation of different excitonic
resonances, it is important to include Coulomb scattering on a microscopic level. For
completeness, we start by investigating the scattering for computation of absorption
spectra, i.e.~for the microscopic polarizations $\ev{P_k} = \ev{v^{\dagger}_k c_k }$.

In the language of the cluster expansion, Coulomb scattering is provided by coupling of
$n$-point expectation values to $(n+2)$-point expectation values. In the case of
the microscopic polarization, we find
%
\bea
\ihddt \ev{P_k} \Bigr|_{\mrm{scatt}} & = & 
\phantom{+}
\sum_{k',q} V_q
\left(
\Delta\ev{v^{\dagger}_{k} c^{\dagger}_{k'} c_{k'+q} c_{k-q}}
-
\Delta\ev{v^{\dagger}_{k+q} c^{\dagger}_{k'} c_{k'+q} c_{k}}
\right)
\nonumber\\
&&
+
\sum_{k',q} V_q
\left(
\Delta\ev{v^{\dagger}_{k} v^{\dagger}_{k'} v_{k'+q} c_{k-q}}
-
\Delta\ev{v^{\dagger}_{k+q} v^{\dagger}_{k'} v_{k'+q} c_{k}}
\right).
\label{eq:ddtP_scat}
\eea
%
This equation is still exact, and the level of approximation is determined by the precision
with which we calculate the four-point correlations. Their dynamics is determined by
an equation of the general form
%
\be
\ihddt \Delta\ev{4} = S[\ev{2}] + T[\ev{2},\Delta\ev{4}] + V[\Delta\ev{6}],
\label{eq:gen4point}
\ee
%
where the four-point correlations are driven by a source term $S[\ev{2}]$ containing
only sums and products of two point terms (such as microscopic carrier occupations and 
polarizations), they are coupled to other four-point correlations in a complicated
manner symbolized by $T[\ev{2},\Delta\ev{4}]$ and they are themselves coupled to
yet higher order terms due to the hierarchy problem. The scattering approximation
to this equation is obtained if all coupling between four-point terms and to six-point
terms is neglected and only the kinetic energy contribution and the two-point source
term is kept on the right-hand side. Formally, the scattering solutions can thus be
obtained from
%
\be
\ihddt \Delta\ev{4}\Bigr|_{\mrm{scatt}} = (\tilde\eps - i \delta)\Delta\ev{4} +  S[\ev{2}],
\label{eq:scat4point}
\ee
%
where we have introduced a phenomenological dephasing constant $\delta$ in addition to
the sum of renormalized free-particle energies $\tilde\eps$. 

More specifically, we obtain
%
\be
\ihddt \Delta\ev{v^{\dagger}_{k} c^{\dagger}_{k'} c_{k'+q} c_{k-q}}
= 
\left(
\tilde\eps^e_{k-q} + \tilde\eps^h_{k} + \tilde\eps^e_{k'+q} - \tilde\eps^e_{k'} - i \delta
\right) \Delta\ev{v^{\dagger}_{k} c^{\dagger}_{k'} c_{k'+q} c_{k-q}}
+
S_{k,k',q}
\label{eq:ddtv+c+cc}
\ee
%
with the singlet source given by
%
\bea
S_{k,k',q}
& = &
(V_q - V_{q+k'-k})
\left( f^e_{k'} (1 - f^e_{k'+q})  (1 - f^e_{k-q}) + (1 -f^e_{k'}) f^e_{k'+q} f^e_{k-q} \right)  \ev{P_k}
\nonumber\\
&&
- V_q
\left( f^e_{k'} (1 - f^e_{k'+q})  (1 - f^h_{k}) + (1 -f^e_{k'}) f^e_{k'+q} f^h_{k} \right) \ev{P_{k-q}}
\nonumber\\
&&
+ V_{q+k'-k}
\left( f^e_{k'} (1 - f^h_{k})  (1 - f^e_{k-q}) + (1 -f^e_{k'}) f^h_{k} f^e_{k-q} \right)  \ev{P_{k'+q}}.
\label{eq:def_S}
\eea
%
Here, we have neglected source terms in third order in $\ev{P_k}$ which is suitable for a low intensity
probe pulse. In general, we want to calculate the susceptibility for a specific carrier density
in a pump-probe type experiment relatively long (up to nanoseconds) after the pump pulse. After 
the fast initial scattering, the carrier distributions are known to be changing very slowly and
can thus be assumed constant. In that case, we do not have to apply the Markov and random-phase
approximation as has often been done in the past. Instead, we see that the quantity 
$\Gamma_k(\omega)$ which we introduced in the main part of the paper in Eq.~(\ref{eq:SBE-Pol}), is
nothing but the Fourier transform of the right-hand side of Eq.~(\ref{eq:ddtP_scat}). This quantity 
can be split into two parts, $\Gamma^c_k$ and $\Gamma^v_k $, corresponding to the first and second 
line of Eq.~(\ref{eq:ddtP_scat}), respectively. 

Beginning with the computation of $\Gamma^c_k$, we need the Fourier transform of Eq.~(\ref{eq:ddtv+c+cc})
which can be formally solved exactly via
%
\be
\Delta\ev{v^{\dagger}_{k} c^{\dagger}_{k'} c_{k'+q} c_{k-q}}(\omega)
=
\frac{S_{k,k',q}(\omega)}
{\hbar \omega -\tilde\eps^e_{k-q} - \tilde\eps^h_{k} - \tilde\eps^e_{k'+q} + \tilde\eps^e_{k'} + i \delta}.
\label{eq:v+c+cc_1}
\ee
%
In a similar manner, we obtain
%
\be
\Delta\ev{v^{\dagger}_{k+q} c^{\dagger}_{k'} c_{k'+q} c_{k}}(\omega)
=
\frac{S_{k+q,k',q}(\omega)}
{\hbar \omega -\tilde\eps^e_{k} - \tilde\eps^h_{k+q} - \tilde\eps^e_{k'+q} + \tilde\eps^e_{k'} + i \delta}.
\label{eq:v+c+cc_2}
\ee
%

With these ingredients, we can calculate $\Gamma^c_k$ by summing over Eqs.~(\ref{eq:v+c+cc_1}) 
and~(\ref{eq:v+c+cc_2}), using the definition of Eq.~(\ref{eq:def_S}),
%
\bea
\Gamma^{c}_{k} & = & 
\phantom{+}
\sum_{k',q} 
\frac{V_{k-q}}
{\hbar \omega -\tilde\eps^e_{q} - \tilde\eps^h_{k} - \tilde\eps^e_{k'} + \tilde\eps^e_{k'+q-k} + i \delta}
\times
\nonumber\\
&&
\Bigl\{
(V_{k-q} - V_{k'-k})
\left( f^e_{k'+q-k} (1 - f^e_{k'})  (1 - f^e_{q}) + (1 -f^e_{k'+q-k}) f^e_{k'} f^e_{q} \right) \ev{P_k}
\nonumber\\
&&
- V_{k-q}
\left( f^e_{k'+q-k} (1 - f^e_{k'})  (1 - f^h_{k}) + (1 -f^e_{k'+q-k}) f^e_{k'} f^h_{k} \right) \ev{P_{q}}
\nonumber\\
&&
+ V_{k'-k}
\left( f^e_{k'+q-k} (1 - f^h_{k})  (1 - f^e_{q}) + (1 -f^e_{k'+q-k}) f^h_{k} f^e_{q} \right) \ev{P_{k'}}
\Bigr\}
\nonumber\\
&&
-
\sum_{k',q} 
\frac{V_{k-q}}
{\hbar \omega -\tilde\eps^e_{k} - \tilde\eps^h_{q} - \tilde\eps^e_{k'} + \tilde\eps^e_{k'+k-q} + i \delta}
\times
\nonumber\\
&&
\Bigl\{
(V_{q-k} - V_{k'-q})
\left( f^e_{k'+k-q} (1 - f^e_{k'})  (1 - f^e_{k}) + (1 -f^e_{k'+k-q}) f^e_{k'} f^e_{k} \right) \ev{P_{q}}
\nonumber\\
&&
- V_{k-q}
\left( f^e_{k'+k-q} (1 - f^e_{k'})  (1 - f^h_{q}) + (1 -f^e_{k'+k-q}) f^e_{k'} f^h_{q} \right) \ev{P_{k}}
\nonumber\\
&&
+ V_{k'-q}
\left( f^e_{k'+k-q} (1 - f^h_{q})  (1 - f^e_{k}) + (1 -f^e_{k'+k-q}) f^h_{q} f^e_{k} \right) \ev{P_{k'}}
\Bigr\}
\label{eq:P_scat_1}
\eea
%
The other quantity $\Gamma^{v}_{k}$ can be obtained by symmmetry considerations.
By exchaning each electron and hole index, we find
%
\bea
\Gamma^{v}_{k} & = & 
\phantom{+}
\sum_{k',q} 
\frac{V_{k-q}}
{\hbar \omega -\tilde\eps^h_{q} - \tilde\eps^e_{k} - \tilde\eps^h_{k'} + \tilde\eps^h_{k'+q-k} + i \delta}
\times
\nonumber\\
&&
\Bigl\{
(V_{k-q} - V_{k'-k})
\left( f^h_{k'+q-k} (1 - f^h_{k'})  (1 - f^h_{q}) + (1 -f^h_{k'+q-k}) f^h_{k'} f^h_{q} \right) \ev{P_k}
\nonumber\\
&&
- V_{k-q}
\left( f^h_{k'+q-k} (1 - f^h_{k'})  (1 - f^e_{k}) + (1 -f^h_{k'+q-k}) f^h_{k'} f^e_{k} \right) \ev{P_{q}}
\nonumber\\
&&
+ V_{k'-k}
\left( f^h_{k'+q-k} (1 - f^e_{k})  (1 - f^h_{q}) + (1 -f^h_{k'+q-k}) f^e_{k} f^h_{q} \right) \ev{P_{k'}}
\Bigr\}
\nonumber\\
&&
-
\sum_{k',q} 
\frac{V_{k-q}}
{\hbar \omega -\tilde\eps^h_{k} - \tilde\eps^e_{q} - \tilde\eps^h_{k'} + \tilde\eps^h_{k'+k-q} + i \delta}
\times
\nonumber\\
&&
\Bigl\{
(V_{q-k} - V_{k'-q})
\left( f^h_{k'+k-q} (1 - f^h_{k'})  (1 - f^h_{k}) + (1 -f^h_{k'+k-q}) f^h_{k'} f^h_{k} \right) \ev{P_{q}}
\nonumber\\
&&
- V_{k-q}
\left( f^h_{k'+k-q} (1 - f^h_{k'})  (1 - f^e_{q}) + (1 -f^h_{k'+k-q}) f^h_{k'} f^e_{q} \right) \ev{P_{k}}
\nonumber\\
&&
+ V_{k'-q}
\left( f^h_{k'+k-q} (1 - f^e_{q})  (1 - f^h_{k}) + (1 -f^h_{k'+k-q}) f^e_{q} f^h_{k} \right) \ev{P_{k'}}
\Bigr\}.
\label{eq:P_scat_2}
\eea
%

The previous two equations can be rewritten in terms of microscopic scattering matrices 
$\gamma^{\mrm{D}}_{k}(\omega)$ and $\gamma^{\mrm{OD}}_{k,k'}(\omega)$ which couple microscopic polarizations
with different $k$-indices with one another. The scattering contribution to the polarization dynamics 
is then given by 
%
\be
\Gamma_{k} =
- i \gamma^{\mrm{D}}_{k}(\omega) \ev{P_k} + i \sum_{k'} \gamma^{\mrm{OD}}_{k,k'}(\omega) \ev{P_{k'}} ,
\label{eq:def_gD_gOD}
\ee
%
where the scattering matrices can be defined by comparison to Eq.~(\ref{eq:P_scat_1}) and~(\ref{eq:P_scat_2}).
While the coupling of the different components of the microscopic polarization leads to a redistribution among 
the $\ev{P_k}$ and thus to dephasing, we can also see from Eqs.~(\ref{eq:P_scat_1})--(\ref{eq:P_scat_2})
that
%
\be
\ihddt \sum_k \ev{P_k} \Bigr|_{\mrm{scatt}} = 0.
\label{eq:sum_Pk}
\ee
%
In terms of the scattering matrices, this relation is guaranteed due to their property
%
\be
\gamma^{\mrm{D}}_{k}(\omega) = \sum_{k'} \gamma^{\mrm{OD}}_{k',k}(\omega).
\label{eq:sum_gOD}
\ee
%
We want to emphasize that for stationary densities and a weak probe pulse it is natural to solve the
equations using the Fourier transformation. In that way, we obtain frequency dependent scattering
matrices where $\omega$ enters in the energy denominator. This is different from the usual
procedure of using Markov and random-phase approximation in the time-domain. We can recover the
old Markov result if for each term we replace the true frequency $\omega$ by the single particle 
frequency $\omega_S$ which describes the main evolution of the respective source.

\subsection{Scattering for photon-assisted polarizations}
\label{app:scat_Pi}
%
For photon-assisted polarizations, the proceeding is quite similar to the previous subsection.
In this case, the photon-assisted two-point quantities couple to photon-assisted four point 
quantities in the form
%
\bea
\ihddt \Delta\ev{b^{\dagger}_q P_k } \Bigr|_{\mrm{scatt}} & = & 
\phantom{+}
\sum_{k',q} V_q
\left(
\Delta\ev{b^{\dagger}_q v^{\dagger}_{k} c^{\dagger}_{k'} c_{k'+q} c_{k-q}}
-
\Delta\ev{b^{\dagger}_q v^{\dagger}_{k+q} c^{\dagger}_{k'} c_{k'+q} c_{k}}
\right)
\nonumber\\
&&
+
\sum_{k',q} V_q
\left(
\Delta\ev{b^{\dagger}_q v^{\dagger}_{k} v^{\dagger}_{k'} v_{k'+q} c_{k-q}}
-
\Delta\ev{b^{\dagger}_q v^{\dagger}_{k+q} v^{\dagger}_{k'} v_{k'+q} c_{k}}
\right).
\label{eq:b+P_scat}
\eea
%
As before, one has to set up equations of motion for these photon-assisted four-point terms
which couple to yet higher correlations. In order to obtain a scattering approximation of
the photon-assisted 4-point terms, we only keep the renormalized kinetic energies and the
main sources. These sources are similar as in Eq.~(\ref{eq:def_S}), but due to the
presence of the photon operator many more factorization possibilities can occur. Under fully
incoherent conditions, however, where we require expectation values of coherent quantities
such as $\ev{b^\dagger}$, $\ev{P_k}$ or $\ev{b^\dagger c^\dagger c}$ to vanish, the only
remaining terms have exactly the form of Eq.~(\ref{eq:def_S}) where each polarization has
to be replaced by the corresponding photon assisted polarization.

In the case of PL, we expect a steady-state solution since the source term of
PL changes adiabatically and can be assumed constant. Nevertheless, the photon
operator in each term provides a frequency dependence via the free rotation of the
photon operator via
%
\be
\ihddt \Delta\ev{b^\dagger_q \hat{O}}\Bigr|_{\mrm{em}} = -\hbar\omega_q \Delta\ev{b^\dagger_q \hat{O}}
\label{eq:ddt_bq}
\ee
%
for each carrier operator $\hat{O}$. When we set the left-hand side to zero, we find the same frequency 
dependence as for the Fourier transformation of the polarization equation.

Consequently, the scattering matrices for photon-assisted polarizations are identical to 
those for the microscopic polarization and the scattering denoted by $T_{k,q}$ in Eq.~(\ref{eq:SLE_BY})
in the main text can be expressed as
%
\be
T_{k,q} =
- i \gamma^{\mrm{D}}_{k}(\omega_q) \ev{b^\dagger_q P_k} 
+ i \sum_{k'} \gamma^{\mrm{OD}}_{k,k'}(\omega_q) \ev{b^\dagger_q P_{k'}} ,
\label{eq:b+P_scat2}
\ee
%
in full analogy to Eq.~(\ref{eq:def_gD_gOD}).

\subsection{Scattering for excitonic correlations}
\label{app:scat_cX}
%
For the calculation of PL we need to know the excitonic correlations within the radiative
cone, i.e. with infinitesimal center of mass momentum. In the main text they are called
%
\be
c_X^{0,k',k} = \Delta\ev{c^{\dagger}_{k} v^{\dagger}_{k'} c_{k'} v_{k} } 
             = \Delta\ev{c^{\dagger}_{k} v_{k} v^{\dagger}_{k'} c_{k'} }
             = \Delta\ev{P^{\dagger}_{k} P_{k'}}.
\label{eq:def_cX0}
\ee
%
At this point, we can briefly address different approximations for the excitonic correlations.
If we completely neglect excitonic correlations in the PL source term of Eq.~(\ref{eq:PL}),
we recover the old Hartree-Fock result \cite{Kira:98}. If instead we set up the equations of
motion for the excitonic correlations in the form of Eq.~(\ref{eq:gen4point}) and treat them
on  a scattering level by neglecting all coupling to four- and six-point terms, we can formally
solve for $c_X$ and express the correlations as functional of carrier densities. In the main text
this is refered to as {\em scattering approximation} of the excitonic source terms. Since photon
number operators and photon-assisted correlations themselves are four-point operators, both
approximations are fundamentally inconsistent since they neglect other four-point terms. This can
be seen as negative PL at the continuum (scattering approximation) or even at higher 
bound states (Hartree-Fock), but nevertheless a qualitatively correct picture is obtained around
the 1s-resonance.

In a consistent approximation, all four-point terms have to be treated on the same footing. 
Thus, the two-point sources and the four-point terms of Eq.~(\ref{eq:gen4point}) have to be
solved and approximations are only introduced on the next level, for the six-point quantities
describing scattering of excitonic correlations. Thus, we have to set up equations of motion
for six-point quantities of the form
%
\be
\ihddt \Delta\ev{6}\Bigr|_{\mrm{scatt}} = (\tilde\eps - i \delta)\Delta\ev{6} +  S[\ev{2},\Delta\ev{4}],
\label{eq:scat6point}
\ee
%
where we keep the free-particle energy and the source term consisting of products of two and four-point
expectation values. Its precise form results from factorizing eight-point terms as e.g.
$\ev{c^{\dagger}_{k} v^{\dagger}_{k'} c^{\dagger}_{k''} c^{\dagger}_{l} c_{l+q'} c_{k''+q-q'} c_{k'} v_{k-q}  }$.
It turns out that all factorized sources contain at least one four-point correlation. 
Due to the indistinguishability of the quantized carriers, the general factorization procedure analogous 
to the previous two subsections leads to a huge variety of different triplet scattering terms. In particular,
carrier operators belonging to one polarization operator in Eq.~(\ref{eq:def_cX0}) can be split into
two different expecation values. The full Coulomb scattering including all those source terms includes
all quantum effects due to the Fermionic character of the excitons as well as possible momentum scattering
between different excitons. This full scattering scenario is currently not feasible in two dimensions. 
Therefore, we make the assumption that scattering acts only within one of the polarization operators of
Eq.~(\ref{eq:def_cX0}). In analogy to the previous section of photon-assisted polarizations, we assume
that the second polarization operator remains unchanged and is not split into different expectation
values. Under this assumption, the derivation of the scattering proceeds analogously as before, the 
number of scattering terms is also identical, and the only major difference is the lack of a frequency
dependence of excitonic correlations. Thus, we obtain exciton scattering of the form
%
\bea
\ihddt c_X^{0,k',k} 
&=& 
 -i \left(
        \left[ \tilde\gamma^{\mrm{D}}_{k} \right]^\star 
             + \tilde\gamma^{\mrm{D}}_{k'}
    \right) c_X^{0,k',k} 
\nonumber\\
&&  +i 
\sum_{k''} 
    \left(
        \left[\tilde\gamma^{\mrm{OD}}_{k,k''}\right]^\star c_X^{0,k',k''} 
            + \tilde\gamma^{\mrm{D}}_{k',k''} c_X^{0,k'',k} 
    \right).
\label{eq:scat_cX}
\eea
%

%

\end{document}